\documentclass[%
 reprint,
 amsmath,amssymb,
 aps,
superscriptaddress,
]{revtex4-2}

\usepackage{graphicx}
\usepackage{dcolumn}
\usepackage{bm}
\usepackage{hyperref}
\hypersetup{colorlinks=true,citecolor=blue,linkcolor=blue,urlcolor=blue}

\usepackage{xparse}
\usepackage[caption=false]{subfig}

\usepackage{appendix}
\usepackage{graphicx}
\usepackage{dcolumn}
\usepackage{bm}
\usepackage[cspex,bbgreekl]{mathbbol}

\renewcommand{\d}[2]{\ensuremath{\frac{\text{d} #1}{\text{d} #2}}}

\newcommand{\ket}[1]{\ensuremath{\left| #1 \right>}}

\newcommand{\braket}[2]{\ensuremath{\left< #1 \ \vphantom{#2} \right| 
\left. #2 \vphantom{#1} \right>}}

\newcommand{\matrixel}[3]{\ensuremath{\left< #1 \vphantom{#3} \right| #2 
\left| #3 \vphantom{#1} \right>}}

\usepackage{bbold}

\usepackage{xparse}
\usepackage{xcolor}
\usepackage{appendix}
\usepackage{graphicx}
\usepackage{dcolumn}
\usepackage{bm}

\newcommand{\magenta}[1]{{\color[cmyk]{0,.9,0,0.2}{#1}}}

\usepackage{bbold}

\newcommand{\re}[1]{(\ref{#1})}

\newcommand {\dis}{\displaystyle}

\newcommand{\beg}{\begin{equation}}
\newcommand{\en}{\end{equation}}

\newcommand{\eps}{\varepsilon}

\renewcommand{\eqref}[1]{Eq.~(\ref{#1})}

\renewcommand{\vec}[1]{ \bm{#1} }

\def\p{{\bm p}}

\def\a {{c}}
\def\ad { c^\dagger}

\def\k {{\bm k}}
\def\q {{\bm q}}

\newcommand{\vex}[1]{\bm{#1}}

\begin{document}
 \title{
Loschmidt Echo of Far-From-Equilibrium Fermionic Superfluids
}

\author{Colin Rylands}
 \email{crylands@umd.edu}
\affiliation{Joint Quantum Institute and
 Condensed Matter Theory Center, University of Maryland, College Park, MD 20742, USA}
\author{Emil A. Yuzbashyan}
 \affiliation{Department of Physics and Astronomy, Center for Materials Theory, Rutgers University, Piscataway, NJ 08854 USA}
  \author{Victor Gurarie}
 \affiliation{Department of Physics and Center for Theory of Quantum Matter,
University of Colorado, Boulder, Colorado 80309, USA}
 \author{Aidan Zabalo}
 \affiliation{Department of Physics and Astronomy, Center for Materials Theory, Rutgers University, Piscataway, NJ 08854 USA}

\author{Victor Galitski}
\affiliation{Joint Quantum Institute and
	Condensed Matter Theory Center, University of Maryland, College Park, MD 20742, USA}

\date{
    \today
}

\date{\today}
\begin{abstract}
Non-analyticities in the logarithm of the Loschmidt echo, known as dynamical quantum phase transitions [DQPTs], are a recently introduced attempt to classify the myriad of possible phenomena which can occur in far from equilibrium closed quantum systems.  In this work, we analytically investigate the Loschmidt echo in  nonequilibrium $s$-wave and topological $p_x+ip_y$ fermionic superfluids.    We find that the presence of non-analyticities in the echo is not invariant under global rotations of the superfluid phase. We remedy this deficiency by introducing a more general notion of a grand canonical Loschmidt echo. Overall, 
our study shows that DQPTs are
not a good indicator for the long time dynamics of an interacting system. In particular, there are no DQPTs  to tell apart   distinct   dynamical phases of quenched  BCS superconductors.    Nevertheless,  they   can signal  a quench induced change
in the  topology   and also keep track of solitons   emerging from unstable stationary states of a BCS superconductor.  
\end{abstract}

\maketitle
\section{Introduction}
Equilibrium phase transitions [EPTs], both classical and quantum, are by now quite well understood due to the existence of a unified theoretical framework which describes their physics. At the center of this is the
partition function which completely characterizes a system. In the thermodynamic limit, the logarithm of the partition function [the free energy] may exhibit non-analytic behavior as a function of temperature [or other system parameter] signifying a phase transition. This behavior is reflected by the system's observables which are given by derivatives of the free energy and can, therefore, be non-analytic themselves. 

Away from equilibrium the situation is less straightforward. For closed quantum systems which are far from equilibrium the natural object to study is the time evolution operator, $U(t)$. This object however, can defy calculation even in non-interacting systems as it depends on both the Hamiltonian of the system and how the system was taken out of equilibrium. 
It has been proposed that a simpler quantity to study is the Loschmidt echo; defined as $\mathcal{L}(t)=|\mathcal{G}(t)|^2$ where~\cite{HeylPolKeh, Heylreview}
\begin{eqnarray}
\mathcal{G}(t)&=&\matrixel{\Psi_i}{  U(t) }{\Psi_i}=\langle \Psi_i |\Psi(t)\rangle,
\label{echo}
\end{eqnarray}
with $\ket{\Psi_i}$ the initial state of the system. $G(t)$ resembles a boundary partition function and likewise its logarithm may exhibit non-analytic points as a function of $t$~\cite{CardCal}. By analogy with EPTs these are called dynamical quantum phase transitions [DQPTs] and by now have been studied in many systems \cite{TrapinHeyl, KhatunBhattacharjee,VajnaBalazs, LangFrankHal,HeylScaling, SharmaSuzukiDutta, KarraschSchuricht, KennesSchurichtKarrasch, Fogarty, VoskAltman, GurarieDQPTS, HalimehYegGurar, BudichHeyl, HeylBudich, BhattacharyaBandDutta, BhattacharyaDutta, CanoviWernerEck,Jafari1,Jafari2,poz1, poz2,poz3,PerfettoPiroliGambassi, RylandsMTM, RylandsAndreiLLwork, AndSirk, LackiHeyl, Zunkovich2018}, most commonly for the particular nonequilibrium situation of a quantum quench~\cite{PolRev, MitraRev, CazalillaJstat,CalabreseCardyJstat, CauxJstat, EsslerFagottiJstat, RylandsAndreiARCMP}. While this non-analytic behaviour as function of time is certainly of interest, the relevance of DQPTs  for the dynamics of a system and in particular  its observables is less obvious than in the equilibrium case. Indeed, observables cannot be expressed as derivatives of $G(t)$. It has been shown however, most notably in the Ising model~\cite{HeylPolKeh} that the period of oscillations of the order parameter coincides with the period of DQPTs in certain quenches. Furthermore, it was seen that for the long range Ising model,  a phase diagram mapped out by the presence of DQPTs coincides with one mapped out by the long time dynamics of the system's order parameter~\cite{Zunkovich2018} suggesting a correspondence between these two notions of dynamical phases.

Much of the information regarding DQPTs has been garnered by analytic studies in free models or through numerical analysis.
In this work we carry out an analytic study of the Loschmidt echo in an interacting, experimentally relevant system -- the nonequilibrium $s$-wave Bardeen-Cooper-Schrieffer [BCS] superconductor -- as well as in the topological $p_x+ip_y$ superfluid. We compare this to the behavior of an analogous noninteracting system and uncover several features which are a consequence of the interactions. We find 
that in an interacting model DQPTs do not occur periodically in time as they do in noninteracting models and moreover,  are often transient. Crucially, DQPTs  fail entirely to signal distinct steady states in the quench dynamics of the BCS superconductor.  This shows
 that \textit{DQPTs are not a reliable indicator of the long time behaviour of an interacting system.}
 
 It will be shown below  that the presence of DQPTs is not invariant under  global  rotations  of the superfluid phase.
 Such rotations do not affect ordinary physical observables of an isolated superfluid, but can induce or altogether remove DQPTs. To resolve this problem, we introduce the notion of \textit{grand canonical Loschmidt echo} $\mathcal{L}_\mu(t)=|\mathcal{G}_\mu(t)|^2$, 
\begin{eqnarray}
\mathcal{G}_\mu(t)&=&\matrixel{\Psi_i}{  U_\mu(t) }{\Psi_i}.
\label{echomu}
\end{eqnarray}
Here $U_\mu(t)$ is the time ordered exponent of $\hat H(t) -\mu(t)\hat N$, $\hat H(t)$ is the system Hamiltonian and $\hat N$ is the total particle number operator.  With a proper choice of $\mu(t)$ DQPTs not only emerge, but also distinguish quenches across quantum critical points from other types of quenches. The $s$-wave BCS dynamics we address in this paper \magenta{are} particle-hole symmetric, which
ensures $\mu(t)\equiv 0$. In this case, the grand canonical echo reduces to the \textit{canonical Loschmidt echo} defined in~\eqref{echo}. On the other hand, for the nonequilibrium $p_x+ip_y$ superfluid $\mu(t)\ne0$.

We explore the significance and meaning of DQPTs for two different nonequilibrium scenarios, the quench dynamics of the ground state following a sudden change in interaction strength~\cite{YDGF,FDGY} and the solitonic dynamics which emerges from a range of unstable stationary states~\cite{Yuzbash}. The  appearance of DQPTs is then compared to the behaviour of the  system at long times which has been well studied previously~\cite{BarakovSpivakLevitov,YuzbashAltKuzEnol,YuzbashTsyplyAltshuler,YuzbashKuzAlt,BarankovLevitov, YuzbashDzero,BarankovLevitov2, PhysRevB.79.132504, Gurarie, Dzero_2009,   Yuzbash,YDGF, FDGY, PhysRevLettFGDY,Scaramazza}. Throughout this paper  \textit{dynamical} or \textit{nonequilibrium phases} are understood as qualitatively distinct long time states of the system distinguished  by  qualitatively different behaviours of the order parameter.

For the quench dynamics of $s$-wave fermionic superfluids, we find that DQPTs \textit{cannot} be used to determine the dynamical phase diagram.
We show that DQPTs are absent throughout the phase diagram except asymptotically when the initial state is the free Fermi gas ground state. This is similar to the analogous non-interacting case -- DQPTs can only occur when the initial state is the normal ground state. In the presence of interactions however, the quenched system is far richer and significantly, the Loschmidt echo is shown to be completely insensitive to the distinct dynamical phases exhibited by the superfluid. Specifically,  it has been established that the long time dynamics following a quench of the interaction strength can be classified into three nonequilibrium phases wherein, at long times, the order parameter amplitude  vanishes [Phase I], approaches a constant value [Phase II] or persistently oscillates [Phase III]. The transitions between these phases are continuous   and occur as one varies the initial and final interaction strengths. No signature of these phase transitions is seen using DQPTs.  

For the soliton dynamics however, we identify an interesting relationship between the number of solitons which are present in the dynamics of the order parameter $\Delta(t)$ and the number of DQPTs. There are two kinds of solitons in the time-dependent BCS problem -- normal and anomalous. Normal solitons emerge from   eigenstates of a free Fermi gas [normal states]. These states display an odd number, $2k-1$, of discontinuities  in the fermion occupation factor  and produce up to $k$ solitons in the order parameter. We find that each  soliton can be associated with a single  DQPT.  A similar, although weaker relationship is also shown to exist for the anomalous solitons when the initial state is an excited stationary state of the BCS model with nonzero gap and $2k$ discontinuities. Here there are two varieties of anomalous solitons and only one of them produces a DQPT. 

Generally we find that the only necessary [though not sufficient] condition for the existence of DQPTs is the presence of zeros in the
distribution function  of Cooper pairs $\gamma[\xi]$. This distribution measures the fraction of pairs  in an instantaneous excited state as a function of the energy.  In the noninteracting system the zeros of $\gamma[\xi]$ are integrals of motion, but can emerge, move around, and disappear altogether in the course of evolution in the interacting case. Thus interactions can potentially remove or induce DQPTs. The transient character of  DQPTs means that  time translation of the initial state may lead to DQPTs being avoided or encountered.  Since the same asymptotic state may be reached from an infinite number of initial states related by time translation but DQPTs may be absent in some of these, this undermines the correlation between DQPTs  and long time dynamics.

In the presence of particle-hole symmetry there is a class of initial states that always produce DQPTs. These are states where the imaginary part of the equal time anomalous Green's function vanishes at the Fermi energy. This property is equivalent to the existence of a
permanent   zero in the Cooper pair distribution  at the Fermi level. Consider the superconducting order parameter integrated 
from initial to present time, $\Phi(t, t_i)=2\int_{t_i}^t dt\Delta(t)$, and taken $\mod 2\pi$, i.e., on the unit circle. A DQPT for these states occurs each
time $\Phi(t)$ crosses $\pi$. In particular, the normal $k$-soliton solutions belong to this class and for them $\Phi(\infty, -\infty)=
2\pi k$ meaning there are exactly $k$ DQPTs for these solutions.

An interesting situation with DQPTs arises in one of the nonequilibrium phases of a topological 2D $p$-wave superfluid.
  The quench phase diagram of this system consists of the same three nonequilibrium phases I, II and III described above, but they are further subdivided into regions of different nonequilibrium topology \cite{FDGY}. We find that there are no DQPTs when using the canonical Loschmidt echo despite the existence of zeros in the Cooper pair distribution. However, a proper choice of $\mu(t)$ in the
  grand canonical echo [such that the phase of the order parameter is time independent],  brings  about   DQPTs in certain regions of the phase diagram. Then, using the known relation between the parity of the number of zeros  of the Cooper pair distribution and nonequilibrium topology, we show that the  number of DQPTs  in the grand canonical echo signals whether Majorana edge modes emerged or disappeared as a result of the quench.  At the same time it tells us whether the quench was across the quantum critical point or not.

This paper is organized as follows: In sections~\ref{II} and \ref{III} we introduce the $s$-wave BCS Hamiltonian, the nonequilibrium problems we are concerned with and the types of initial states that shall be considered. In section~\ref{nointsect} we outline the quench dynamics of the analogous non-interacting problem with a view to later compare this with our results for the interacting model. We highlight several features which are not present in the interacting case. In section~\ref{laxsec} we review the main method which is used to study our systems; an approach based on  classical integrability. In  section~\ref{gssec} we derive the Loschmidt echo for the system when quenched from the ground state under a change in interaction strength. In section~\ref{solitonsec} we discuss the soliton dynamics emanating from unstable stationary states. We derive analytic expressions for the Loschmidt echo and express them in terms of the time dependent order parameter. It is shown that DQPTs occur in conjunction with the appearance of zeros in the Cooper pair distribution. In the subsequent section we explore the connection with the Cooper pair distribution for some deliberately engineered initial conditions using  numerical simulation of the system. In section~\ref{pip} we  study DQPTs in a related model, the 2D $p_x+ip_y$ superfluid, and in the penultimate section we propose the notion of the grand canonical Loschmidt echo.  In the final section we summarize our results as well as discuss open questions and the meaning of DQPTs beyond the main focus of this work on their relation to the nonequilibrium dynamics of  interacting systems.

\section{Hamiltonian and Loschmidt Echo} 
\label{II}

The Hamiltonian of the $s$-wave BCS model is given by
\begin{eqnarray}\label{H}
\hat H(g)=\sum_{\vec{p}\sigma}\xi_{\vec{p}}c^\dag_{\vec{p}\sigma}c_{\vec{p}\sigma}-g\sum_{\vec{p},\vec{q}}c^\dag_{\vec{p}\uparrow}c^\dag_{-\vec{p}\downarrow} c_{-\vec{q}\downarrow} c_{\vec{q}\uparrow},
\end{eqnarray}
where $c^\dag_{\vec{p},\sigma}, c_{\vec{p},\sigma}$ are creation and annihilation operators for fermions with spin $\sigma=\uparrow, \downarrow$ and momentum $\vec{p}$ and $\xi_{\vec{p}}$ are the corresponding single particle energy levels relative to the Fermi level.  Fermions interact via a pairing interaction of strength $g$.   The model separates into decoupled sectors wherein each level is singly occupied, called the blocked sector, or either empty or doubly occupied called the unblocked sector.  We shall consider here the case where there are no states in the blocked sector, levels are either empty or doubly occupied.

The Hamiltonian is quantum integrable~\cite{RICHARDSON, Gaudin} and can be solved via Bethe Ansatz. However, since it contains infinite range interactions, the mean field description becomes exact in the thermodynamic limit \cite{doi:10.1063/1.523493,Roman:2002dh,PhysRevB.71.094505} and provides a simpler approach to the system. 
This remains true even out of equilibrium \cite{Faribault_2009,Wu_unpublished} and therefore we have that at time $t$ the system is in the state
\begin{eqnarray}\label{psit}
\ket{\Psi_\mathrm{BCS}(t)}=\prod_{\vec p}\left[u^*_{\vec{p}}(t)+v^*_{\vec{p}}(t)c^\dag_{\vec{p}\uparrow}c^\dag_{-\vec{p}\downarrow}\right]\ket{0}. 
\end{eqnarray}
Here $\ket{0}$ is the vacuum which contains no particles and $z^*$ denotes the complex conjugate of $z$. The coefficients $u_{\vec{p}}(t)$ and $v_{\vec{p}}(t)$ are solutions of the   Bogoliubov-de-Gennes [BdG] equations,
\begin{eqnarray}\label{BdG}
i\partial_t\begin{pmatrix}u_{\vec{p}}\\
v_{\vec{p}}
\end{pmatrix}=\begin{pmatrix} \xi_{\vec{p}}& \Delta(t)\\
\Delta^*(t)& -\xi_{\vec{p}}
\end{pmatrix}\begin{pmatrix}u_{\vec{p}}\\
v_{\vec{p}}
\end{pmatrix},
\end{eqnarray}
which follow from~\eqref{psit} and the mean field form of the Hamiltonian
\beg
\hat H(t) =\sum_{\vec{p}\sigma}\xi_{\vec{p}}c^\dag_{\vec{p}\sigma}c_{\vec{p},\sigma}-\Biggl(\Delta(t)\sum_{\vec{p}}c^\dag_{\vec{p}\uparrow}c^\dag_{-\vec{p}\downarrow} +\mathrm{h.c.}\Biggr).
\label{mfH}
\en
Here $\Delta(t)$ is the time dependent superconducting order parameter defined as
\begin{equation}\label{SCcondition}
\begin{split}
\Delta(t)=g \sum_p\matrixel{\Psi_\mathrm{BCS}(t)}{ c_{-\vec{p}\downarrow} c_{\vec{p}\uparrow} }{\Psi_\mathrm{BCS}(t)}\\
=g\sum_{p}u_{\vec{p}}(t)v_{\vec{p}}^*(t).
\end{split}
\end{equation}
This is the self-consistency condition for the mean field approach and needs to be solved in conjunction with the BdG equations.  

The simplicity of the state~\re{psit} means that the Loschmidt echo can be readily evaluated, 
\begin{eqnarray}\label{Gt}
\mathcal{L}(t)=\prod_{\vec{p}}\left|u_{\vec{p}}^*(0)u_{\vec{p}}(t)+v_{\vec{p}}^*(0)v_{\vec{p}}(t)\right|^2\!\!,
\end{eqnarray}
where the initial state is encoded in the initial conditions $u_{\vec{p}}(0), v_{\vec{p}}(0)$. A DQPT will occur when at some time the Loschmidt Echo vanishes, meaning that the time evolved state, $\ket{\Psi_\mathrm{BCS}(t)}$ momentarily becomes orthogonal to its initial value. This translates to there being a $\tau$ and $\vec{q}$ such that $u_{\vec{q}}(\tau)/v_{\vec{q}}(\tau)=-v^*_{\vec{q}}(0)/u_{\vec{q}}^*(0)$.

The evolution of the system may be more conveniently analyzed using classical Anderson pseudospins~\cite{Anderson} which are defined as 
\begin{equation}
\begin{split}
&2s^z_{\vec{p}}=\langle\hat n_{\vec{p}}\rangle-1=|v_{\vec{p}}|^2-|u_{\vec{p}}|^2,\\ 
&s^-_{\vec{p}}=\langle c_{-\vec{p}\downarrow} c_{\vec{p}\uparrow}\rangle=u_{\vec{p}}v_{\vec{p}}^*,\quad s^+_{\vec{p}}=\left(s^-_{\vec{p}}\right)^*\!\!,
\end{split}
\label{pseudospins}
\end{equation}
where $\hat n_{\vec{p}}$ is the total occupation number operator for states $|\vec p\uparrow\rangle$ and $|-\vec p \downarrow\rangle$, $s^\pm_{\vec{p}}$ define the spin components $s^x_{\vec{p}}, s^y_{\vec{p}}$ through
$s^\pm_{\vec{p}}=s^x_{\vec{p}}\pm i s^y_{\vec{p}}$, and quantum averages are with respect 
to the time dependent wavefunction of the system $|\Psi_\mathrm{BCS}(t)\rangle$. Note that the spin length $|\vec{s}_{\vec{p}}|=1/2$.

 In terms of $\vec{s}_{\vec{p}}$, the BdG equations become the Bloch equations   for a system of spins evolving in a time dependent magnetic field  
\begin{eqnarray}\label{spinEOM}
\dot{\vec{s}}_{\vec{p}}=\vec{B}_{\vec{p}}\times \vec{s}_{\vec{p}},
\end{eqnarray}
where $\vec{B}_{\vec{p}}=[-2\Delta_x(t), -2\Delta_y(t), 2\xi_{\vec{p}}]$, $\Delta_x(t)$ and $-\Delta_y(t)$ are the real and imaginary parts of the  order parameter $\Delta(t)\equiv\Delta_x(t)-i\Delta_y(t)$, and the self consistency condition is 
\begin{equation}
\Delta(t)=g\sum_{\vec{p}} s^-_{\vec{p}}.
\label{Deltaspins}
\end{equation}
Note that if in the initial state  $\vec{s}_{\vec{p}}$ depend on $\vec{p}$ only through $\xi_{\vec{p}}$, as is the case for all initial conditions we consider in this paper, then this remains true throughout 
the  time evolution and we can write $\vec{s}_{\vec{p}}(t)=\vec{s}(\xi_{\vec{p}}, t)$.

Let us also introduce the Cooper pair distribution function $\gamma_\p$ which is the cosine of the angle between the spin
$\vec{s}_{\vec{p}}$ and its magnetic field $\vec{B}_\p$,
\begin{equation}
\label{CPDF}
\gamma_\p\equiv\gamma[\xi_{\vec{p}}, t]\equiv\cos{[\theta_{\vec p}]}=\frac{2{\vec{s}_{\vec{p}}} \cdot\vec{B}_{\vec{p}}}{|\vec{B}_{\vec{p}}|}.
\end{equation}
In particular, $\gamma[\xi_{\vec{p}}, t_\perp]=0$   means $\vec{s}_{\vec{p}}$ is perpendicular to $\vec{B}_{\vec{p}}$  at $t=t_\perp$. 
The general solution of the BdG equations [the Cooper pair wavefunction]  in terms of $\gamma_\p$ is
 \beg
 \begin{pmatrix}u_{\vec{p}}\\
v_{\vec{p}}
\end{pmatrix}=\sqrt{ \frac{1+\gamma_\p}{2} } e^{-i E_\p^+ t} \varphi_\p^+ + \sqrt{ \frac{1-\gamma_\p}{2} } e^{-i E_\p^- t} \varphi_\p^-,
\en
where $E_\p^\pm=\pm\sqrt{\xi_\p^2+|\Delta(t)|^2}$ and $\varphi_\p^\pm$ are the instantaneous eigenvalues and eigenstates of
the BdG Hamiltonian [the $2\times2$ matrix in~\eqref{BdG}].  In equilibrium $\gamma_\p$ and $\varphi_\p^\pm$ are time independent
and $\gamma_\p$ takes values $\gamma_\p=\pm1$ indicating that the energy level $\xi_{\vec{p}}$ is occupied by a ground state $(-1)$ or an excited $(1)$   Cooper pair~\cite{BCS}. In the pseudospin language $\gamma_\p=1 (-1)$ corresponds to the spin being parallel (antiparallel) to its magnetic field. Out of equilibrium  $\gamma_\p$ is generally  time dependent, can take any  value in the interval $[-1, 1]$, and determines the instantaneous probability distributions of the two states $\varphi_\p^\pm$  of a Cooper pair with   instantaneous energies $E_\p^\pm$.

At weak coupling the pairing is confined to a narrow energy window around the Fermi energy. Then the density of states is constant and for every single-particle energy level $\xi_{\vec{p}}$ there is a corresponding level at $-\xi_{\vec{p}}$. Under these conditions, the $s$-wave BCS Hamiltonian~\re{H}  is invariant under  a particle-hole transformation 
\beg
c_{\vec{p}\sigma}\to c^\dag_{\vec{p}\sigma},\quad 
c^\dag_{\vec{p}\sigma}\to c_{\vec{p}\sigma},\quad \xi_\p\to -\xi_p.
\en
 This symmetry is unbroken in the BCS ground state. 
In the language of classical pseudospins~\re{pseudospins},  a state is particle-hole symmetric  when
\begin{equation}
s^x(\xi_{\vec{p}})=s^x(-\xi_{\vec{p}}),\quad s^{y,z}(\xi_{\vec{p}})=-s^{y,z}(-\xi_{\vec{p}}),
\label{phsym}
\end{equation}
where $\vec{s}(\xi_{\vec{p}})\equiv \vec{s}_{\vec{p}}$.  These relations are preserved by the equations of motion~\re{spinEOM}, i.e., they  hold at all times
when they hold for the initial state. 
We shall only consider such particle-hole symmetric initial states for the $s$-wave superconductor in this work. \eqref{phsym} implies $\Delta_y(t)\equiv 0$, $\Delta(t)=\Delta_x(t)$  and therefore
\begin{equation}
\vec{B}_{\vec{p}}=-2\Delta(t)\hat x+ 2\xi_{\vec{p}}\hat z,
\label{Bph}
\end{equation}
where $\hat x$ and $\hat z$ are the unit vectors along the $x$ and $z$-axis, respectively. Note also that  the Cooper pair distribution
$\gamma[\xi_\p, t]$ is an even function of $\xi_\p$ in this case.

 The Loschmidt echo can be compactly expressed in the spin language as
\begin{eqnarray}\label{SpinEcho}
\mathcal{L}(t)=\prod_{\vec{p}}\left[\frac{1}{2}+2\vec{s}_{\vec{p}}(0)\cdot\vec{s}_{\vec{p}}(t)\right],
\end{eqnarray}
 and a DQPT can now be seen to occur when there is a spin $\vec{s}_{\vec{q}}(t)$ which is flipped relative to $\vec{s}_{\vec{q}}(0)$. 
 
 Our mean-field description is valid only in the thermodynamic limit wherein  the mean  spacing of single particle levels $\delta\to0$. After we have obtained expressions for $\mathcal{L}(t)$ we shall take the thermodynamic limit in the $s$-wave model via first changing the product over levels to a sum 
 \begin{equation}
 \prod_{\vec{p}}f(\xi_{\vec{p}})=\exp{\sum_{\vec{p}} \text{ln}{[f(\xi_{\vec{p}})}]},
 \end{equation}
  for any function $f$.  After which we take 
  \begin{equation}
  \sum_{\vec{p}}\to \nu V\int_{-D}^D d\xi \,\nu(\xi),
  \end{equation}
where  $V$ is the volume of the system, $2D$ is the bandwidth and $\nu$ is the density of states.  We work in the infinite bandwidth limit, $D\to \infty$. 

\section{nonequilibrium protocols and initial states}
\label{III}

We shall be primarily concerned with two different nonequilibrium scenarios. One is a sudden quench of the interaction strength, $g_i\to g_f$ with the initial state taken to be the ground state of \magenta{$\hat{H}(g_i)$}.
Such a nonequilibrium protocol  has been extensively studied previously in this and related models~\cite{BarakovSpivakLevitov,YuzbashAltKuzEnol,YuzbashTsyplyAltshuler,YuzbashKuzAlt,BarankovLevitov, YuzbashDzero,BarankovLevitov2, PhysRevB.79.132504, Gurarie, Dzero_2009,   Yuzbash,YDGF, FDGY, PhysRevLettFGDY,Scaramazza}. The dynamics can classified into three distinct phases characterized by the long time behavior of the order parameter which either vanishes [Phase I], approaches a constant [Phase II] or persistently oscillates [Phase III]. Phases I and III have no analogue in a non-interacting system and emerge due to the interactions which impose the time dependent self consistency condition on the order parameter given in \eqref{SCcondition}. Furthermore this nonlinear constraint allows for non-trivial dynamics to emerge without a quench when perturbing an unstable stationary state~\cite{Yuzbash}, this is the other nonequilibrium scenario we shall consider. Unstable stationary states  are unstable equilibria of classical equations of motion~\re{spinEOM} similar to an inverted pendulum, but generally with more dynamical degrees of freedom. Upon perturbing away from them the system can be classified by the resulting solitonic behaviour of the order parameter. 

The types of initial states that we will consider can be grouped into two categories, anomalous and normal. Anomalous initial states are described by the spin distributions  
\begin{eqnarray}\label{AnomSpin}
2s^x_{\vec{p}}(0)=\frac{-e_{\vec{p}}\Delta_\mathrm{in}}{\sqrt{\xi_{\vec{p}}^2+\Delta_\mathrm{in}^2}},~2s^z_{\vec{p}}(0)=\frac{-e_{\vec{p}}\xi_{\vec{p}}}{\sqrt{\xi_{\vec{p}}^2+\Delta_\mathrm{in}^2}},
\end{eqnarray}
and $s^y_{\vec{p}}(0)=0$, where $e_{\vec{p}}=\pm 1$ and $\Delta_\mathrm{in}=g_i\sum_{\vec{p}} s^x_{\vec{p}}(0)$. Depending on the choice of $e_{\vec{p}}$ this will correspond to either the ground state wherein $e_{\vec{p}}=1$ for all $\vec{p}$ or some excited state. 
The value of $\Delta_\mathrm{in}$ in the ground state we denote $\Delta_i$.
The excited states we consider are particle-hole symmetric and consist of flipping a number of spins symmetrically about the Fermi level at $\xi_F=0$, e.g., the choice $e_{\vec{p}}=\text{sgn}[|\xi_{\vec{p}}|-a]$ excites the quasiparticles in the region $\xi_{\vec{p}}\in [-a,a]$ about the Fermi level.  Anomalous excited states naturally have an even number of discontinuities in their spin distribution. 

The normal states are eigenstates of $\hat{H}(g_i)$ which are simultaneously eigenstates of the free Fermi gas, i.e., of $\hat H(0)$. Their spin distribution is given by 
\begin{eqnarray}\label{NormSpin}
s_{\vec{p}}^z(0)=\frac{e_{\vec{p}}}{2},~s^x_{\vec{p}}(0)=s^y_{\vec{p}}(0)=0,
\end{eqnarray}
where again the choice of $e_{\vec{p}}=\pm 1$ determines whether this is the ground state or an excited state of the free gas, e.g., the ground sate is described by $s^z_{\vec{p}}(0)=-\text{sgn}(\xi_{\vec{p}})/2$ and $e_{\vec{p}}=-\text{sgn}[\xi_{\vec{p}}(\xi_{\vec{p}}^2-a^2)]$ excites the particles in the region $\xi_{\vec{p}}\in [-a,a]$ about the Fermi level. Normal states contain an odd number of discontinuities.

\section{Non-interacting Quenches}
\label{nointsect}

Before studying the full model described by  Eqs.~\re{BdG} and \re{SCcondition} we briefly recall how the system behaves in the analogous noninteracting quench, when the self consistency condition~\re{SCcondition} is not enforced. The quench is then characterized by a change of a constant pairing potential in the mean field Hamiltonian~\re{mfH}; $\hat{H}(\Delta_\mathrm{in})\to \hat{H}(\Delta_\mathrm{fn})$ rather than a change in the pairing strength, $g_i\to g_f$.  In this case the dynamics are still described by \eqref{BdG} however now in these equations $\Delta(t)\to\Delta_\mathrm{BdG}(t)=\Delta_\mathrm{in}\Theta(-t)+\Delta_\mathrm{fn}\Theta(t)$,    which is unrelated to $\Delta(t)$ in~\eqref{SCcondition}. Here $\Theta(t)$ is the Heaviside function. Such non-interacting dynamics can be efficiently solved by finding the canonical transformation which relates the eigenstates  of the pre and post quench Hamiltonian~\cite{Iucci}. 

For an initial state described by the spin distribution~\re{AnomSpin} the Loschmidt echo is found to be~\cite{RylandsMTM}
\begin{eqnarray}\label{NonintEcho}
\mathcal{L}(t)&=&\mathcal{F}\prod_{\vec{p}}\Big|1+\tan^2\!{\left[\frac{\theta_{\vec{p}}}{2}\right]}e^{i e_{\vec{p}}2E^\mathrm{fn} _{\vec{p}}t}\Big|^2,\\\nonumber
\tan{\!\left[\frac{\theta_{\vec{p}}}{2}\right]\!}&=&\!\frac{\sqrt{ \mathcal{E}^+_{\vec{p}}(\Delta_\mathrm{fn})  \mathcal{E}^+_{\vec{p}}(\Delta_\mathrm{in})}- \sqrt{ \mathcal{E}^-_{\vec{p}}(\Delta_\mathrm{fn})  \mathcal{E}^-_{\vec{p}}(\Delta_\mathrm{in})}}{ \sqrt{ \mathcal{E}^-_{\vec{p}}(\Delta_\mathrm{fn})  \mathcal{E}^+_{\vec{p}}(\Delta_\mathrm{in})}+\sqrt{ \mathcal{E}^+_{\vec{p}}(\Delta_\mathrm{fn})  \mathcal{E}^-_{\vec{p}}(\Delta_\mathrm{in})} },
\end{eqnarray}
where $\mathcal{F}$ is a time independent constant, $E^\mathrm{fn}_{\vec{q}}=\sqrt{\xi_{\vec{p}}^2+\Delta_\mathrm{fn}^2}$ is the energy of a quasiparticle of the post quench Hamiltonian,  $\mathcal{E}^\pm_{\vec{p}}(\Delta)=\sqrt{\xi_{\vec{p}}^2+\Delta^2}\pm\xi_\p$ and $\theta_{\vec{p}}/2$ is the angle of rotation for the BdG transformation which relates $\hat{H}(\Delta_\mathrm{in})$ and $\hat{H}(\Delta_\mathrm{fn})$.  This expression is typical of quenches between quadratic, fermionic Hamiltonians and exhibits several features which are common to all such situations. By inspecting \eqref{NonintEcho} we see that DQPTs may only occur if there exists a $\q$ such that $|\!\tan{[\theta_{\vec{q}}/2]}|=1$, in which case they occur periodically with the period, $T_\text{DQPT}=\pi/E^\mathrm{fn}_{\vec{p}}$ depending only on the final Hamiltonian parameters through $\Delta_\mathrm{fn}$. Furthermore, the angle $\theta_{\vec{p}}$ is independent of the choice of $e_{\vec{p}}$  and therefore, DQPTs for non-interacting systems are insensitive to the particular eigenstate of $H(\Delta_\mathrm{in})$ which is taken to be the initial state, $\ket{\Psi_i}$. We shall see below that when interactions are included both the existence of DQPTs and their period depends upon the choice of   $e_{\vec{p}}$ as well as $g_i$ and $g_f$. 

In addition to these general properties, we can also note some aspects which are specific to the present scenario. The condition for a DQPT to occur can only be satisfied if $\Delta_\mathrm{in}=0$ i.e, a quench from a normal eigenstate. If this is the case then as a consequence of particle-hole symmetry it is the spin at the Fermi level, $\xi_F=0$, which is flipped relative to its initial position resulting in a DQPT with period $T_\text{DQPT}=\pi/\Delta_\mathrm{fn}$. The dynamics of the order parameter~\re{SCcondition} after a quench of this type are straightforward to evaluate~\cite{Iucci}.   If the initial state is the normal ground state,  then at long time the order parameter approaches  a constant and exhibits damped oscillations with period which coincides with $T_\text{DQPT}$,
\begin{eqnarray}
 \Delta(t)\approx \Delta_\mathrm{fn} (1-\lambda^*\ln 2)+\lambda^* \frac{\sin[2\Delta_\mathrm{fn} t]}{2 t},
\end{eqnarray}
where $\lambda^*= \ln^{-1}(2D/\Delta_\mathrm{fn})$.   We note that unlike the self-consistent dynamics we study in subsequent sections, this answer is somewhat pathological  within the standard theory of superconductivity, which is  applicable only in the weak  coupling limit $D\to\infty$.
Similar behaviour also occurs when $\Delta_\mathrm{in}>0$ however, as mentioned above, in that case no DQPTs occur.  As we shall discuss below,  the dynamics of the order parameter are markedly different when interactions are present.

Further insight can be gained by presenting these results in the language of spins. The dynamics are still described by \eqref{spinEOM} but with $\vec{B}_{\vec{p}}(t)=\vec{B}^\mathrm{in}_{\vec{p}}\Theta(-t)+\vec{B}^\mathrm{fn}_{\vec{p}}\Theta(t)$ where $\vec{B}^\mathrm{in/fn}_{\vec{p}}=(-2\Delta_\mathrm{in/fn},0,-2\xi_{\vec{p}})$ and so the system consists of a collection of decoupled spins each precessing around its own constant magnetic field. The angle of rotation for the Bogoliubov transformation can then be interpreted as the angle between the final magnetic field and the initial spins.
 A DQPT  occurs if there exists a spin which is perpendicular to the magnetic field i.e. the Cooper pair distribution function~\re{CPDF} has a zero, $\gamma[\xi_{\vec{q}}]=0$ for some $\xi_{\vec{q}}$. In this case the spin rotates in the plane perpendicular to the field allowing it to become flipped relative to its initial position. Moreover, since the magnetic field is independent of time, this occurs periodically with period $T_\text{DQPT}=2\pi/|\vec{B}_{\vec{q}}^\mathrm{fn}|$.  There may exist multiple spins with vanishing Cooper pair distribution function in which case many DQPTs will exist each with its own period. 

An important point to note here is that the DQPTs discussed in this section are permanent in the following sense: The zeros of $\gamma[\xi_{\vec{p}}]$ are constants of motion and so at any point in the evolution the  condition $\gamma[\xi_{\vec{q}}]=0$ is satisfied if that is the case initially. Accordingly $\ln|\braket{\Psi(t_0)}{\Psi(t)}|$ shall exhibit the exact same non-analytic behavior for arbitrary $t_0$ not just for $t_0=0$ and do so with the same period. DQPTs cannot, therefore, be removed by simply translating our initial state in time.  This recurrent behavior is significant if one is interested in the long time dynamics of a system. For example, a quench from a certain initial state may result in the appearance of DQPTs and some behaviour of its observables at long time. An infinite number of initial states related by time translation give rise to the same long time dynamics and observables and owing to their recurrent nature the same DQPTs are present also. 

The spin interpretation provides some intuition of what can be expected in the presence of interactions. In that case the magnetic field is not constant in time but evolves along with the system and  a spin that is initially orthogonal to the magnetic field may not remain so. More precisely, the zeros of the now time dependent Cooper pair distribution  can be transient and may appear or disappear as a function of time. This allows for the possibility that interactions remove or induce DQPTs when compared to the noninteracting system. Moreover in the presence of interactions, DQPTs in general will not  occur periodically and as a result of this may be avoided through time translation of the initial state.  Such transient behavior would then remove any connection between the long time dynamics and DQPTs. 


\section{Lax Vector}
\label{laxsec}

The nonequilibrium dynamics of the BCS model, including the self consistency condition~\re{SCcondition}, have been extensively studied and the behavior of $\Delta(t)$ and spin distribution $\vec{s}_{\vec{p}}(t)$ for many initial states determined. The methods by which this has been achieved are naturally more complicated than those of the previous section and  central to them is the special
connection    between  integrals of motion and the long time dynamics of the system, which emerges for pairing Hamiltonians in the thermodynamic limit. In this section we briefly review this method and how it is applied to the  BCS model. We refer the reader to \cite{YDGF} and \cite{YuzbashKuzAlt} for further details.  

The equations of motion in the spin representation (\ref{spinEOM}) are equivalent to those of a system of classical spins governed by the Hamiltonian, 
\begin{eqnarray}
H_\text{c}=\sum_{\vec{p}}2\xi_{\vec{p}}s^z_{\vec{p}}-g_f\sum_{{\vec{p}},\vec{q}}s^+_{\vec{p}}s^-_{\vec{q}},
\end{eqnarray} 
where the  spins obey the Poisson bracket $\{s_{\vec{p}}^i,s_{\vec{q}}^j\}=-\epsilon_{ijk}\delta_{{\vec{p}},{\vec{q}}}s^k$. This classical Hamiltonian is Liouville integrable meaning that we can construct $n$ functionally independent integrals of motion in 
involution~\cite{Arnold1989}, where $n$ is the number of degrees of freedom [the number of distinct energy levels $\xi_{\vec{p}}$ each of which is to represent one classical spin $\vec{s}(\xi_\p)$]. The integrability follows from the 
Lax representation~\cite{Lax1968,babelon_bernard_talon_2003} of the equations of motion~(\ref{spinEOM})
\begin{equation}
\frac{dL}{dt}= i[M, L],
\label{lax}
\end{equation}
where $L$ and $M$ (the Lax pair) are the following two $2\times 2$ matrices:
\begin{equation}
L=
\begin{pmatrix}
L_z(u) & L_-(u)\\
L_+(u) & -L_z(u)\\
\end{pmatrix},
\quad
M=\begin{pmatrix}
u & \Delta \\
\Delta^* & -u\\
\end{pmatrix},
\end{equation}
$u$ is an auxiliary (spectral) parameter, $L_\pm(u)\equiv L_x(u)\pm iL_y(u)$  and $L_x(u)$, $L_y(u)$, and $L_z(u)$ are the components
of the \textit{Lax vector},
\begin{eqnarray}\label{Lax}
\vec{L}(u)=-\frac{\hat{z}}{g_f}+\sum_{ \vec{p}} \frac{\vec{s}_{\vec{p}}}{u-\xi_{\vec{p}}}.
\end{eqnarray}
It is more convenient to work with the Lax vector $\vec{L}(u)$ than the Lax matrix $L$. In terms of the Lax vector, the Lax equation~(\ref{lax}) becomes
\begin{equation}
\dot{ \vec{L} }(u) =\vec{B}(u)\times \vec{L}(u),
\end{equation}
where $\vec{B}(u)=[-2\Delta_x(t), -2\Delta_y(t), 2u]$.
This equation implies that the length of this vector is conserved, for arbitrary values of $u$, under evolution with $H_c$ i.e.,
\begin{eqnarray}
\d{\vec{L}^2(u)}{t}=0.
\end{eqnarray}
Thus it can be evaluated for an initial spin distribution using \eqref{AnomSpin} or \eqref{NormSpin} after which it must remain a constant  and serves as a generator for the integrals of motion of the system. Specifically, the residues of $\vec{L}^2(u)$ at simple poles at 
$u=\xi_{\vec p}$ provide the $n$ independent integrals of motion thus proving the Liouville integrability.

 Zeros of  $\vec{L}^2(u)$  are also integrals of motion and are especially useful for understanding the dynamics.
 After bringing \eqref{Lax} to a common denominator, it is not too difficult to see that the square of the Lax vector may be expressed as 
\begin{eqnarray}
\vec{L}^2(u)=\frac{Q_{2n}(u)}{g^2_f\prod_{\xi_{\vec{p}} } (u-\xi_{\vec{p}})},
\end{eqnarray}
where $Q_{2n}(u)$ is a degree $2n$ polynomial known as the \textit{spectral polynomial} whose roots are the zeros of $\vec{L}^2(u)$.  The dynamics of the system may be completely discerned by knowing the structure of the roots of $Q_{2n}(u)$ which are either real and doubly degenerate or come in complex conjugate pairs. Moreover, from its definition $\vec{L}^2(u)=L_x^2(u)+L_y^2(u)+L^2_z(u)$ one can see that any real zero of $\vec{L}^2(u)$ is also a zero of each of the components $L_{x,y,z}(u)$. For the situations we are considering the roots  will densely fill the real line in the thermodynamic limit apart from a number of isolated complex conjugate roots. Remarkably, the number and pattern of these isolated roots determines the long time behaviour of $\Delta(t)$ and $\vec{s}_{\vec{p}}(t)$.

Just as there are many choices for the integrals of motion there are many choices of dynamical variables in which to analyze the dynamics. A particularly convenient choice is to use the zeros $u_j$  of $L_-(u)=0$,
\beg
L_-(u_j)= \sum_{ \vec{p} } \frac{s^-_{\vec{p}}}{u_j-\xi_{\vec{p}}}=0,
\label{sepv}
\en
in terms of which the equations of motion separate and can be integrated. These are related to the spin variables via
\begin{eqnarray}\label{uspin}
s^-_{\vec{p}}(t)&=&\frac{\Delta(t)}{g_f}\frac{\prod_{j=1}^{n-1}(\xi_{\vec{p}}-u_j)}{\prod_{ \xi_{\vec{p}} \neq \xi_{\vec{q}} }(\xi_{\vec{p}}-\xi_{\vec{q}})},
\end{eqnarray}
which can be proven directly from the definition of $L_-(u)$ and \eqref{Lax}. 
The advantage of using the $u_j$ as the dynamical variables instead of the spins can be seen by examining their equations of motion which are,
\begin{eqnarray}\label{uEOM}
\dot{u}_j&=&\frac{2i \sqrt{Q_{2n}(u_j)}}{\prod_{k\neq j}(u_k-u_j)},
\\\label{DeltaEOM}
\dot{\Delta}(t)&=&2i\Delta(t)\sum_{j=1}^{n-1}u_j,
\end{eqnarray}
where we specialized~\eqref{DeltaEOM} to the particle-hole symmetric case [all other equations and results in this section are general].
From this one can immediately see that if a separation variable $u_j$ coincides with a root of $Q_{2n}(u)$ then its equations of motion are automatically satisfied and furthermore the system of equations is reduced in the number of variables by one. This can be carried out for any number of variables so that if there are $m$ variables which coincide with roots of $Q_{2n}(u)$ then the system is reduced to $n-m-1$ variables satisfying the same set of equations. This is then equivalent to the dynamics of $n-m$ spins which significantly reduces the complexity of the problem. When all separation  variables coincide with roots of $Q_{2n}(u)$ the state is a stationary state of the system.



\section{Ground state Quench Dynamics}
\label{gssec}

Having laid some groundwork we now examine where and when DQPTs occur if the system is quenched, $g_i\to g_f$, from the ground state of $\hat{H}(g_i)$ with $g_i\neq g_f$. We concentrate on the long time behaviour of the system post quench where it is possible to derive analytic expressions for $\Delta(t)$ and $\vec{s}_{\vec{p}}(t)$. Moreover, the short time dynamics is sensitive to microscopic details of the initial state and Hamiltonian with universal behavior only emerging in the long time limit. 

The long time behaviour of the BCS model when quenched from the ground state can be classified into three distinct phases which are characterized by $\Delta(t)$~\cite{YuzbashTsyplyAltshuler,BarankovLevitov,YDGF,YuzbashDzero}. Which phase is realized depends  on a single external control parameter $\beta\equiv \lambda_f^{-1}-\lambda_i^{-1}= \ln[\Delta_i/\Delta_f]$.
Here $\lambda_{i,f}=V\nu g_{i,f}$ are the initial and final dimensionless BCS coupling constants and $\Delta_{i,f}=2D e^{-1/\lambda_{i,f}}$ are the corresponding ground state gaps. The parameter $\beta$ controls  the nature of the isolated roots of $Q_{2n}(u)$. Phase I corresponds to $\beta\geq \pi/2$ and $Q_{2n}(u)$ having no isolated roots. In this phase $\Delta(t)\to 0$ as $t\to\infty$ and the spins $\vec{s}_{\vec{p}}$ in the nonequilibrium steady state precess around the $z$-axis with angular frequencies $2\xi_{\vec{p}}$, while the steady state wavefunction  is a time dependent superposition of normal states. Phase II corresponds to $-\pi/2<\beta<\pi/2$ and $Q_{2n}(u)$ possessing a single pair of complex conjugate roots at $\pm i\Delta_\infty$. In this phase $\Delta(t)\to \Delta_\infty$ as $t\to\infty$, where $0<\Delta_\infty\le \Delta_f$. In the Phase II steady state the spin $\vec{s}_{\vec{p}}$ precesses around   a constant field $\vec{B}_{\vec{p}}=(-2\Delta_\infty, 0, 2\xi_{\vec{p}})$ and the wavefunction is therefore a time dependent superposition of anomalous states. Phase III corresponds to $\beta<-\pi/2$ and $Q_{2n}(u)$ having two pairs of complex conjugate roots at $\pm i(\delta_+\pm\delta_-)/2$. In this phase $\Delta(t)$ exhibits persistent periodic oscillations between the values $\delta_-$ and $\delta_+$ and has a compact analytic form $\Delta(t)=\delta_+\text{dn}\,[\delta_+(t-\tau),k]$ where dn$[x,k]$ is the Jacobi elliptic function of modulus $k=1-(\delta_-/\delta_+)^2$. Contained within this phase is a quench from the ground state of the free Fermi gas. The transition between the different phases is continuous and the spin distribution can be determined by first considering it in Phase III and then taking the limit $\delta_-\to\delta_+=\Delta_\infty$ to enter Phase II and then $\Delta_\infty\to 0$ for Phase I.


\subsection{Normal Initial State}
\label{1normalsoliton}

We begin by first examining the quench from the noninteracting system, $g_i=0$. This  is a quench from a quantum
critical point at $g=0$, which    separates the normal, $g\le0$, and  superconducting,   $g>0$ ground states. 
The initial state is described by \eqref{NormSpin} with the choice $e_{\vec{p}}=-\text{sgn}[\xi_{\vec{p}}]$. 
Evaluating $\vec{L}^2(u)$ for this configuration one finds $Q_{2n}(u)$ has $n-2$ doubly degenerate real roots which in the thermodynamic limit merge with $\xi_{\vec{p}}$ and a pair of doubly degenerate complex conjugate roots $\pm i\Delta_f/2$~\cite{Yuzbash}. The presence of the real roots reduces the problem to that of a single separation variable $u_1(t)$. From \eqref{uEOM} one finds that 
\begin{eqnarray}
u_1(t)=-i\Delta_f\tanh{[\Delta_f(t-\tau)]}/2
\end{eqnarray} 
with $\tau$ a constant of integration which depends on how one perturbs away from the unstable state. The presence of $\tau$ reflects the fact that the Fermi ground state is actually a stationary state of the BCS model, albeit an unstable one [see further discussion below]. Accordingly we should think of the state as being the limit of  a vanishing pairing interaction.   

The spin dynamics and order parameter $\Delta(t)$ arising from this initial state can then subsequently be found via Eqs.~\re{uspin} and \re{DeltaEOM}. They are
\begin{equation}
\begin{split}\label{spin1soliton}
s^z_{\vec{p}}(t)=-\frac{\text{sgn}[\xi_{\vec{p}}]}{2}\left[1-\frac{2\Delta(t)^2}{(2\xi_{\vec{p}})^2+\Delta_f^2}\right]\\
s^-_{\vec{p}}(t)=\text{sgn}[\xi_{\vec{p}}]\frac{2\xi_{\vec{p}}\Delta(t)-i\dot{\Delta}(t)}{(2\xi_{\vec{p}})^2+\Delta_f^2},
\end{split}
\end{equation}
where 
\begin{equation}
\label{NormalDelta}
\Delta(t)=\frac{\Delta_f}{\cosh{(\Delta_f(t-\tau))}}.
\end{equation}
We see that $\Delta(t)$ consists of a single soliton~\cite{BarakovSpivakLevitov}; it interpolates between stationary states at $t-\tau=\pm\infty$ and exhibits a single peak at $t=\tau$ coinciding with the point when $u_1(t)$ crosses the real line. In addition, from \eqref{spin1soliton} we see that at $t=\tau$ spins close to the Fermi level are almost flipped relative to their initial position.   In the thermodynamic limit one can therefore expect a DQPT to occur. 

The Loschmidt echo for the single soliton emerging from the Fermi ground state can then be found using~\eqref{SpinEcho}. It is
\begin{equation}
\begin{split}
\mathcal{L}_\mathrm{1s}(t,\Delta_f, \tau)&=\prod_{\vec{p}}\left[1-\frac{\Delta^2(t)}{(2\xi_{\vec{p}})^2+\Delta_f^2}\right]\\
&\displaystyle =
e^{-\pi \nu V\Delta_f\left[1-\sqrt{ 1- \frac{\Delta(t)^2}{\Delta_f^2} } \right]},
\end{split}
\label{singleEcho}
\end{equation}
where in the second line we have gone to the thermodynamic limit. The echo becomes non-analytic when the argument of the square root vanishes at $t=\tau$. Thus there is a single DQPT which occurs exactly at the peak of the soliton at which point  the order parameter reaches the equilibrium ground state value $\Delta(\tau)=\Delta_f$.  

The non-analytic behaviour of the echo can be investigated by expanding about $t=\tau$ from which one sees that 
\begin{eqnarray}\label{logEcho}
l(t)\equiv-\ln{[\mathcal{L}(t)]/V}\approx\pi\Delta_f \nu\left[1- |t-\tau|\right].
\end{eqnarray} 
This scaling in the neighbourhood of a DQPT is similar to that which occurs in the 1D Ising and related models such as the noninteracting quench discussed above~\cite{HeylScaling}. 

The existence of a DQPT when the system is quenched from the normal ground state is similar to the noninteracting case. In contrast however we see that it does not appear periodically and as we anticipated earlier it is transient. We can translate the initial state forward in time to any $t_0>\tau$ in which case the DQPT and the order parameter peak are avoided but the same long time limit is reached. This can be seen explicitly by focusing on the dynamics of the spins to either side of the discontinuity. In the thermodynamic limit they are perpendicular to the magnetic field and complete a single rotation by $2\pi$ around this during the total evolution of the system. If we take the initial state to be any state described by \eqref{spin1soliton} with $t=t_0\le\tau$, i.e., before the peak of the soliton, then there exists a point in time at $t>\tau$ at which these spins  become antiparallel to their initial orientation.  If however we choose $t_0>\tau$, there is no
such point and the DQPT does not occur.

As mentioned in the Introduction, the normal state is just one member   of a much broader class of particle-hole symmetric initial states that have   a permanent zero in the Cooper pair distribution  at the Fermi energy leading to DQPTs.  
Indeed, consider any particle-hole symmetric initial state where the
$x$-component of spins at the Fermi surface is zero. From~\eqref{pseudospins} we see that $s^-_{\vec{p}}= i F_{\vec{p}}(t, t)$, where $F_{\vec{p}}(t, t)$ is the
anomalous Green's function. Thus $s^x_{\xi_{\vec{p}}=0}=0$ is equivalent to $\mathrm{Im\,} F_{\vec{p}}(t, t)=0$ on the Fermi surface. This means that the spins at $\xi_{\vec{p}}=0^{\pm}$, which we denote as $\vec{s}_\pm$, are perpendicular to the $x$-axis. The magnetic field~\re{Bph} at the Fermi surface is $\vec{B}_0(t)=-2\Delta(t) \hat x$. Since the field is  along the $x$-axis at all times, the spins $\vec{s}_\pm$ rotate around this axis with variable angular velocity $2\Delta(t)$ always remaining perpendicular to it. Therefore, the Cooper pair distribution~\re{CPDF} has a permanent zero at the Fermi level. Near the Fermi level $\gamma[\xi_\p, t]=\kappa(t) |\xi_\p|$. For the soliton one can confirm this directly with the help of Eqs.~\re{spin1soliton} and
 \re{NormalDelta} by taking the limits $\xi_{\vec{p}}\to0^\pm$.

 The condition $s^x_{\xi_{\vec{p}}=0}=0$ coupled with particle-hole symmetry~\re{phsym}
 imply a `strong' discontinuity at the Fermi energy meaning that the jump of in $\vec{s}_{\vec{p}}$ across the Fermi surface is the maximum possible. And vice versa maximum jump $|\Delta\vec{s}_{\vec{p}}|=1$ requires $s^x_{\xi_{\vec{p}}=0}=0$. Thus `maximum
 discontinuity at the Fermi surface' and `$s^x_{\xi_{\vec{p}}=0}=0$' are synonymous and this  is true for any particle-hole symmetric state. Conversely,  less than the maximum discontinuity, $|\Delta\vec{s}_{\vec{p}}|<1$, implies $s^x_{\xi_{\vec{p}}=0}\ne0$, which
 removes the zero in the Cooper pair distribution and associated DQPTs.

 The angle of rotation of spins $\vec{s}_\pm$ around the $x$-axis from their initial positions at time $t_i$ is
 \beg
 \Phi(t, t_i)=2\int_{t_i}^t dt\Delta(t).
 \label{Phi}
 \en
 These spins are flipped with respect to their initial orientations whenever $\Phi(t, t_i)=(2m-1)\pi$ with integer $m$. In particular,
 for the soliton~\re{NormalDelta} we have $\Phi(\infty, -\infty)=2\pi$ and $\Phi(\tau, -\infty)=\pi$, indicating a single DQPT at $t=\tau$.   If we take $t_i>\tau$, there is no
DQPT as the spins   do not have enough time to rotate by $\pi$ despite the permanent zero in the Cooper pair distribution at the Fermi level. This shows that, unlike the non-interacting case, the existence of zeros in this distribution   is not a sufficient  condition for a DQPT to occur. Note also that even though the zero is permanent here,
  \eqref{Phi} implies that DQPTs do not occur periodically except for special $\Delta(t)$. This again is in contrast to the non-interacting case where $\Delta(t)=\mbox{const}$ and DQPTs are always periodic.

Thus the DQPT for this limiting quench is linked to a permanent zero in $\gamma[\xi_{\vec{p}},t]$ which is reminiscent of the noninteracting case.
 For more general initial conditions where $s^x_{\xi_{\vec{p}}=0}\ne0$  and which lead to DQPTs, the  zeros of $\gamma[\xi_{\vec{p}}, t]$ are at $\xi_{\vec{p}}\ne0$. These zeros are not protected by the particle-hole symmetry, because the direction of the field $\vec{B}_{\vec{p}}$ changes in time for $\xi_{\vec{p}}\ne0$, and are time dependent as the result.  They  emerge and disappear
and their locations  in general move in the course of the evolution. Indeed, in many other cases we  study below we find that zeros of $\gamma[\xi_{\vec{p}},t]$ are transient in the presence of interactions. In the interacting case it is not obvious that DQPTs necessarily require a permanent or transient zero. Nevertheless we will find that in all our examples DQPTs are always accompanied by zeros of $\gamma[\xi_{\vec{p}},t]$.

\subsection{Superconducting Initial State}
\label{scinst}

We now consider the solution where the initial state is described by \eqref{AnomSpin} with $\Delta_\mathrm{in}=\Delta_i=2De^{-1/V\nu g_i}$, $g_i\neq 0$ and $e_{\vec{p}}=1$ starting in Phase III.  The solution was derived in~\cite{YDGF} in the following manner; first one can show that the  fact that the spectral polynomial $Q_{2n}(u)$ has two pairs of complex isolated roots for $\beta<-\pi/2$ implies that $\Delta(t)$ asymptotes at large times to
\begin{eqnarray}
\Delta(t)=\delta_+\text{dn}\,[\delta_+(t-\tau),k].
\end{eqnarray} 
The next step is to observe, that there is another solution of the spin equations of motion with the same $\Delta(t)$.  It corresponds to the situation when $Q_{2n}(u)$ has the same two pairs of complex isolated roots, while all remaining roots are real. As discussed in section~\ref{laxsec}, there is only one dynamic separation variable in this situation similar to the single soliton, meaning that the equations of motion can be readily solved. Denoting the spins in this particular solution $\vec{\sigma}_\p$, we find
\begin{equation}
\label{sigmazm}
\begin{split}
\sigma_{\vec{p}}^z(t)=\frac{(2\xi_{\vec{p}})^2+\delta_+^2+\delta_-^2-2\Delta^2(t)}{\sqrt{P_4(\xi_{\vec{p}})}},\\ 
\sigma^-_{\vec{p}}(t)=\frac{-4\xi_{\vec{p}}\Delta(t)-2i\dot{\Delta}(t)}{\sqrt{P_4(\xi_{\vec{p}}) }},
\end{split}
\en
where $P_4(u)=[(2u)^2+(\delta_++\delta_-)^2][(2u)^2+(\delta_+-\delta_-)^2]$. Note that all $\sigma_{\vec{p}}(t)$ are periodic in time with the same period (synchronised), which is the period of $\Delta(t)$.

Then, we determine  the Bogoliubov amplitudes $(U_{\vec{p}},V_{\vec{p}})$ for this solution using the relations $U_\p V_\p^*=\sigma_\p^-$, $|U_\p|^2-|V_\p|^2=2\sigma_\p^z$  and the BdG equations~\re{BdG}.  Importantly,  $U_\p$ and $V_\p$ are \textit{not} synchronized due to on overall $\p$-dependent phase, which cancels in $\sigma_\p^-$ and $\sigma_\p^z$. Finally, we notice that the orthogonal two component wavefunction $(V_{\vec{p}}^*,-U_{\vec{p}}^*)$ is another, linearly independent solution  with the same $\Delta(t)$. It follows that the most general solution   with this $\Delta(t)$ a linear combination of the two
\begin{eqnarray}
\begin{pmatrix}u_{\vec{p}}\\
v_{\vec{p}}\end{pmatrix}=\cos{\left[\frac{\theta_{\vec{p}}}{2}\right]}\begin{pmatrix}U_{\vec{p}}\\
V_{\vec{p}}\end{pmatrix}+\sin{\left[\frac{\theta_{\vec{p}}}{2}\right]}\begin{pmatrix}V^*_{\vec{p}}\\
-U_{\vec{p}}\end{pmatrix},
\end{eqnarray}
where $\theta_{\vec{p}}$, the angle which mixes these two solutions, is determined by calculating the integrals of motion via $\vec{L}^2(u)$ and matching them to the pre-quench initial state. 

In terms of spins we can write this solution as 
\begin{eqnarray}\label{asymptotespin}
\vec{s}_{\vec{p}}(t)=\frac{\cos{\left[\theta_{\vec{p}}\right]} }{2}\vec{\sigma}_{\vec{p}}+ \vec{s}^\perp_{\vec{p}}.
\end{eqnarray}
 The  term $\vec{s}_{\vec{p}}^\perp$  rotates around $\vec{\sigma}_{\vec{p}}$ with angular frequency that disperses with $\xi_{\vec{p}}$
 as a consequence of the $\p$-dependent overall phase of $U_\p$ and $V_\p$. Therefore, the actual asymptotic solution $\vec{s}_\p$ for the quench dynamics contains as many frequencies as there are degrees of freedom (spins), unlike  $\sigma_{\vec{p}}$ which are singly periodic.   It also satisfies the self consistency condition only asymptotically at long time when the integral of $\vec{s}_{\vec{p}}^\perp$ over $\xi_{\vec{p}}$ dephases and can be dropped.  

Using \eqref{asymptotespin} in the Loschmidt echo we have that 
\begin{equation}
\begin{split}
\mathcal{L}(t)&=\prod_{\vec{p}}\left[\frac{1}{2}+\vec{s}_{\vec{p}}(0)\cdot\left(\cos{\left[\theta_{\vec{p}}\right]}\vec{\sigma}_{\vec{p}}+2 \vec{s}^\perp_{\vec{p}}\right)\right]\\
&\approx \prod_{\vec{p}}\left[\frac{1}{2}+\cos{\left[\theta_{\vec{p}}\right]}\vec{s}_{\vec{p}}(0)\cdot\vec{\sigma}_{\vec{p}}\right]\!,
\end{split}
\label{LongTimeEcho}
\end{equation}
where in the second line we have dropped the dispersing term whose contribution vanishes at large times. The accuracy of this approximation shall be verified by comparing our analytic expressions for the echo with numerical simulations.

To evaluate $\mathcal{L}$ we require  $\cos{[\theta_{\vec{p}}]}$. This was  calculated explicitly in \cite{YuzbashDzero} and is given by
\begin{eqnarray}
\cos{[\theta_{\vec{p}}]}=\sum_{\sigma=\pm}\sigma\frac{z(\xi_{\vec{p}})}{i\pi }\left[\sqrt{A_\sigma^2\Delta_i^2+(\xi_{\vec{p}} A_\sigma+\beta)^2}\right]\!\!,
\label{Aexp}
\end{eqnarray}
where 
\begin{equation}
A_\pm=\frac{1}{2\sqrt{\xi_{\vec{p}}^2+\Delta_i^2}}\left[\pm i\pi+\ln{\left(\frac{\xi_{\vec{p}}(\xi_{\vec{p}}+\sqrt{\xi_{\vec{p}}^2+\Delta_i^2})}{\Delta_i(\Delta_i+\sqrt{\xi_{\vec{p}}^2+\Delta_i^2})}\right)}\right]
\end{equation}
and $z(\xi_{\vec{p}})=\pm 1$ with the sign determined by the constraint that $\cos{[\theta_{\vec{p}}]}$ should be smooth and 
 tend to $\mp1$ as $\xi_{\vec{p}}\to \pm\infty$.

The complexity of these expressions means that in order to obtain compact analytic forms for the echo we must take some simplifying limits, which shall nevertheless be indicative of the general behaviour. In particular we shall examine the cases $\Delta_i/\Delta_f\ll 1$ and   $\Delta_i\lesssim\Delta_f$ which lie at the edges of the Phase III region. We begin with the former wherein we can expand to leading order in $\Delta_i$ to get $\cos{[\theta_{\vec{p}}]}=2s^z_{\vec{p}}(0)+\mathcal{O}(\Delta_i^2/|\beta|)$ and also $\delta_+=\Delta_f$ , $\delta_-=2\Delta_i|\beta|$. Substituting these along with  Eq.~\re{sigmazm}   into \eqref{LongTimeEcho}, we find
\begin{equation}
\begin{split}
\mathcal{L}(t)= e^{-\pi \nu V\Delta_f\left(1-\sqrt{1-\frac{ \Delta^2(t)+2\Delta_i\Delta(t)}{\Delta_f^2} } 
 -a_D \right),  }\\
   a_D=\frac{ 2\Delta(t)[2\Delta_i-\Delta(t)]} {\pi D\Delta_f},
\end{split}
\label{eq:LE_p3}
\end{equation}
where we have retained the leading finite bandwidth correction $a_D$, so as to more accurately match numerical simulations and also dropped any terms which are higher order in $\Delta_i/\Delta_f$. 
Evidently, this recovers the expression we found in the previous section upon taking $\Delta_i=0$ and with it the single DQPT which appears.  
When  $\Delta_i\neq 0$ however there are no DQPTs as the argument of the square root never vanishes. This can be confirmed by comparing this formula with numerical simulations for a large number of spins, see  Fig.~\ref{fig:sc_sc_p3}. We also note that \eqref{singleEcho} gives significantly worse agreement with numerics, i.e. each peak  in  Fig.~\ref{fig:sc_sc_p3} cannot be described by a sum of solutions of the form of \eqref{singleEcho}. The agreement improves if we use \eqref{eq:LE_p3} without the finite bandwidth correction and improves even more when this correction is included. 

\begin{figure}[hbt]
    \centering
    \subfloat[$\Delta_i = 0.001, \Delta_f = 1.0$\label{fig:sc_sc_p3}]{\includegraphics[width=0.9\columnwidth]{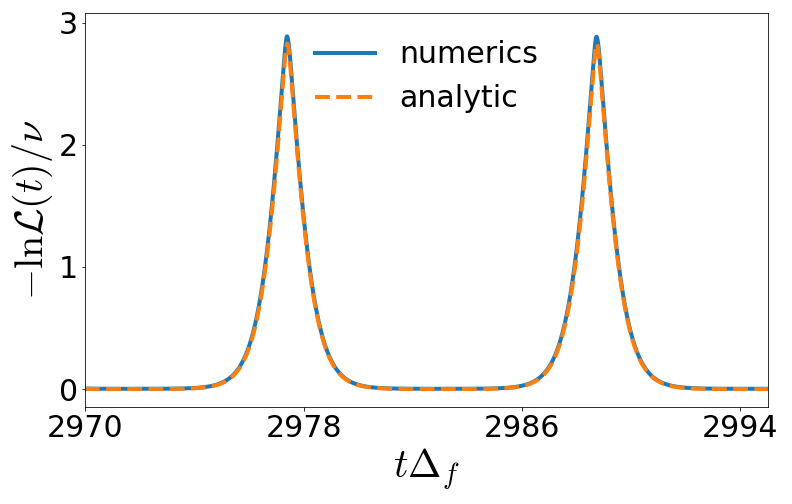}}\\
    \subfloat[$\Delta_i = 0.187, \Delta_f =  1.0$\label{fig:sc_sc_p3p2}]{\includegraphics[width=0.9\columnwidth]{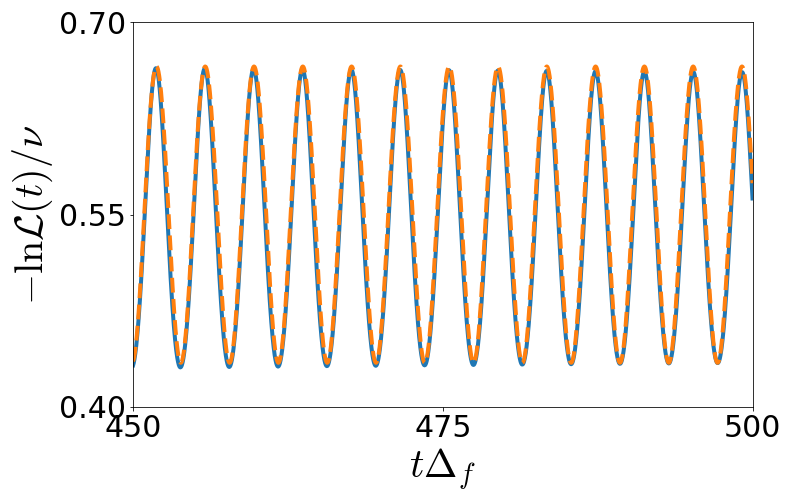}}
    \caption{Log of the Loschmidt echo for interaction quenches in Phase III illustrating  that there are no DQPTs in this phase.    We show two quenches near
    the boundaries of this phase, (a)  for very small $\Delta_i$ and (b) near the Phase II to III transition. Numerical simulations were performed with $n = 50,000$ spins and uniformly spaced single-particle levels $\xi_\p$.   Numerics are compared to the analytic results in \eqref{eq:LE_p3} for (a) and \eqref{BoundaryEcho} for (b). Here and in all other figures interaction quenches are specified by the ground state gaps $\Delta_i$ and $\Delta_f$  for the initial and final interaction strength and we choose half bandwidth $D = 10$,  which sets our energy and time units. }
    \label{fig:sc_sc_p3s}
\end{figure}

\begin{figure}[hbt]
    \centering
    \subfloat[$\Delta_i = 1.0, \Delta_f = 0.001$\label{fig:sc_sc_p1_D}]{\includegraphics[width=.5\columnwidth]{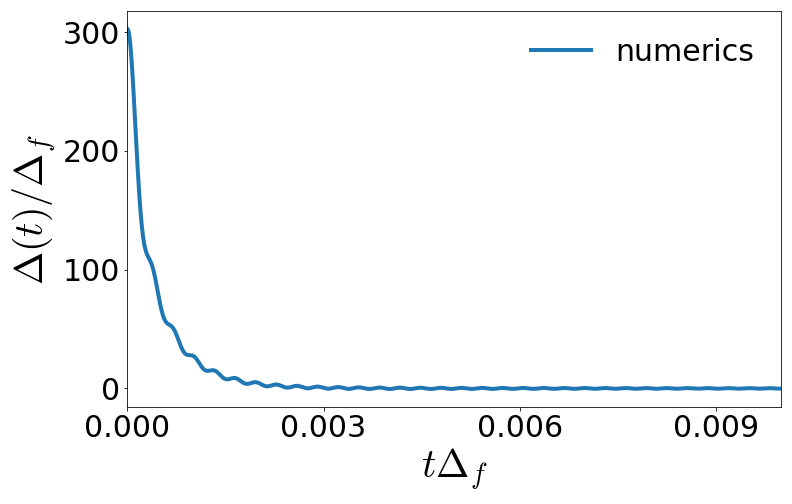}}
    \subfloat[$\Delta_i = 1.0, \Delta_f = 0.001$\label{fig:sc_sc_p1_F}]{\includegraphics[width=.5\columnwidth]{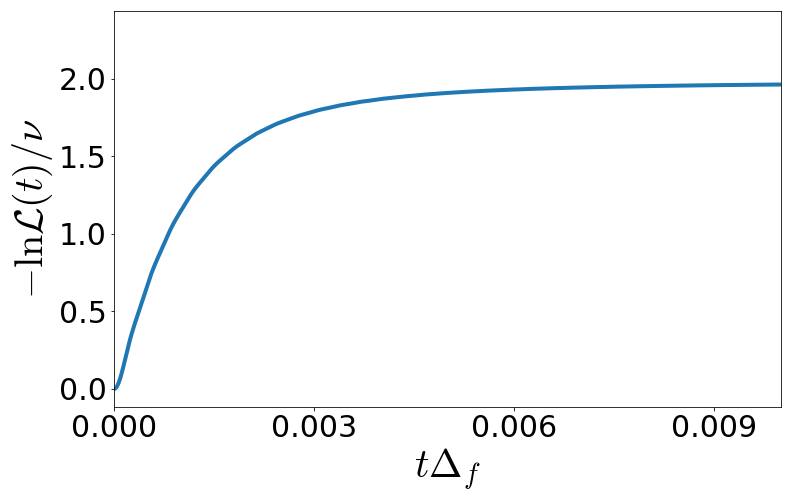}} \\
    \subfloat[$\Delta_i = 0.6, \Delta_f = 0.8$\label{fig:sc_sc_p2_D}]{\includegraphics[width=.5\columnwidth]{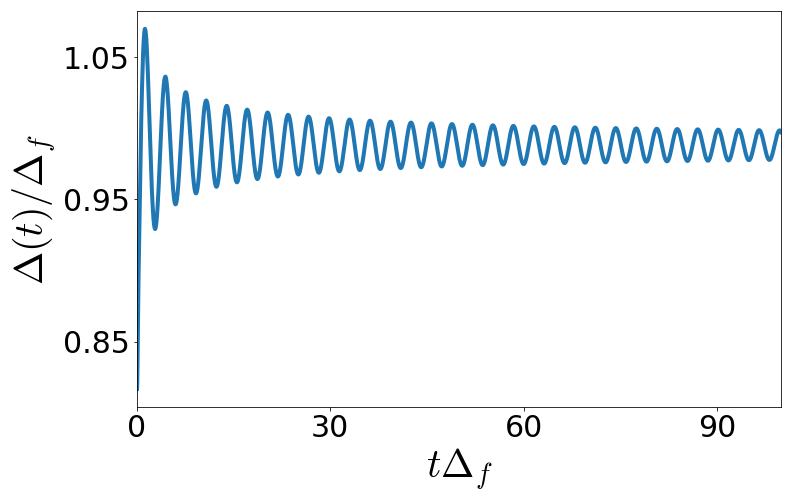}}
    \subfloat[$\Delta_i = 0.6, \Delta_f = 0.8$\label{fig:sc_sc_p2_F}]{\includegraphics[width=.5\columnwidth]{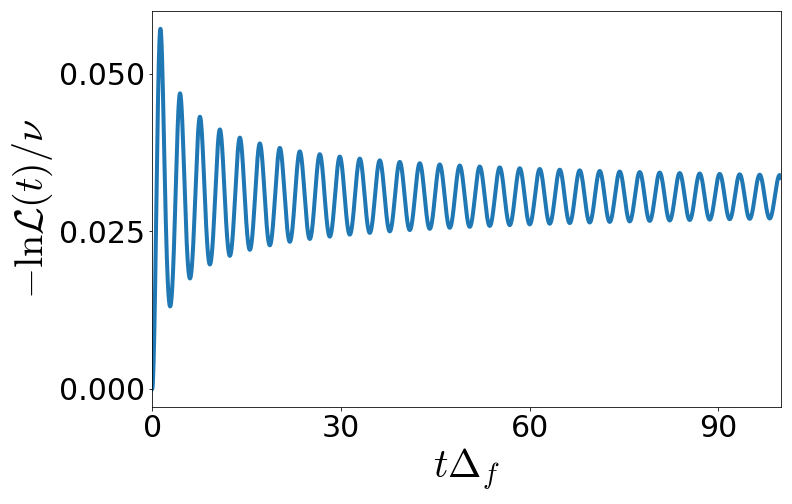}}
    \caption{Order parameter $\Delta(t)$ and  log of the Loschmidt echo for quenches in Phases I and II. In Phase I, at late times (a) the order parameter vanishes and  (b) the  Loschmidt echo approaches a constant. In Phase II,  (c) the  order parameter and  (d) the Loschmidt echo exhibit damped oscillations and decay to a constant. Other parameters are the same as in Fig.~\ref{fig:sc_sc_p3s}. There are no DQPTs in either phase. }
    \label{fig:sc_sc_p1_p2}
\end{figure}

 \begin{figure}[hbt]
    \centering
       \subfloat[$\Delta_i = 0.001, \Delta_f = 1.0$\label{fig:3early}]{\includegraphics[width=.5\columnwidth]{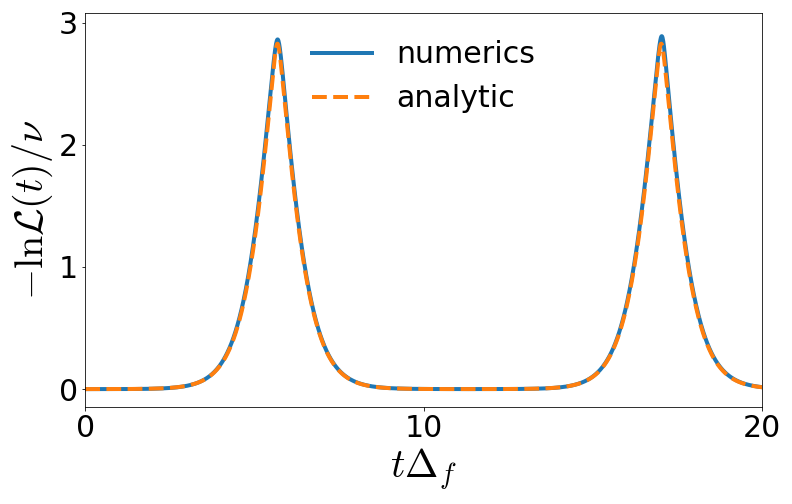}}
         \subfloat[$\Delta_i = 0.6, \Delta_f = 0.8$\label{fig:2early}]{\includegraphics[width=.5\columnwidth]{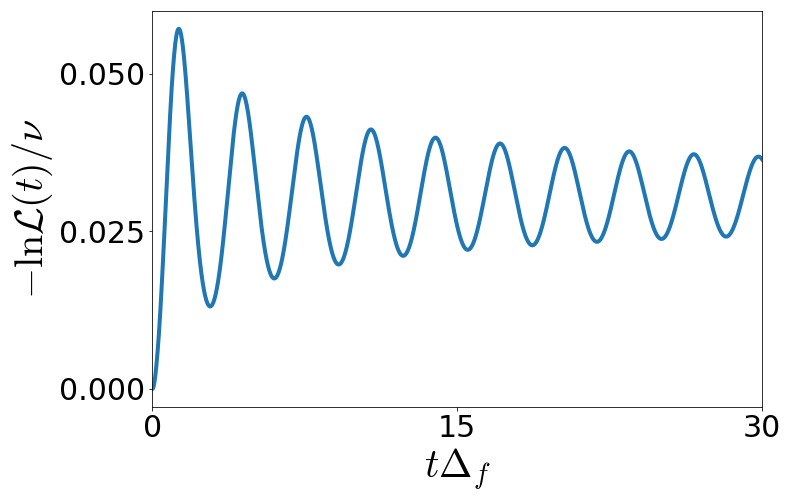}} 
   \caption{Same quenches as in Figs.~\ref{fig:sc_sc_p3} and \ref{fig:sc_sc_p2_F}, but we show the log of the Loschmidt echo at early times. Again there are no DQPTs. Note that~\eqref{eq:LE_p3} [dashed line] we derived for late times  equally well works for early times.}
\label{same_but_early}
\end{figure}
 
Before commenting on this further we examine the alternative limit close to the Phase II and III transition, $\delta_-\lesssim \delta_+$. In this region the order parameter at large times simplifies to 
\begin{eqnarray}
\Delta(t)=\Delta_s\left[1+q\cos{(2\Delta_s t)}\right],
\end{eqnarray}
where $2\Delta_s=\delta_++\delta_-$ and $2q=1-\delta_-/\delta_+\ll 1$. Using these expressions in \eqref{LongTimeEcho} and retaining only the leading terms we have that the  steady state echo is
\begin{eqnarray}\label{BoundaryEcho}
\mathcal{L}(t)=e^{-\alpha_0-\alpha_1 q\cos{(2\Delta_st)}}
\end{eqnarray}
with $\alpha_{0,1}$ being the first and second coefficients in an expansion in $q$. Their explicit form can be determined without too much difficulty but is not necessary for the present discussion. What we have found is that  at long time the Loschmidt echo oscillates with the same period and  in phase with the order parameter, see Fig.~\ref{fig:sc_sc_p3p2}. Additionally the expression is analytic and hence no DQPTs occur. This behaviour is indicative of the whole Phase III region; the echo exhibits persistent oscillations in step with the order parameter and no DQPTs occur unless $\Delta_i\to 0$. 

As mentioned above, the transition between phases is continuous and the behaviour within Phase II can be determined simply from \eqref{BoundaryEcho} by taking $q=0$. The time dependent term then drops out and the  echo becomes a constant depending on $\Delta_i$ and $\Delta_f$. For Phase I the order parameter vanishes at long time. The echo behaves similarly to Phase II approaching a constant at long time which is dependent only on $\Delta_i$ and $\Delta_f$. In both phases no DQPTs occur. The echo and order parameter for quenches within Phase I and II are shown in  Fig.~\ref{fig:sc_sc_p1_p2}. Since our analytical results are based on the known long time steady state, we also numerically checked that there are no DQPTs at early times, see Fig.~\ref{same_but_early}.

The lack of DQPTs when quenching from the ground state of a superconductor can be understood from simple arguments. Earlier, we identified that a DQPT occurs when there exists a spin which becomes flipped relative to itself. For a quench from the normal ground state this was the spin at the Fermi level. The ground state spin distribution of the superconductor is continuous, a property which is preserved by the equations of motion. In addition both the spins at $\xi_{\vec{p}}={\pm\infty}$ and at the Fermi level are static in the post quench system. Thus the post quench system is described by continuous distribution of spins which is pinned at either end, $s_{\xi_{\vec{p}}=\pm\infty}^z=\mp 1/2$, and the Fermi level, $s_{\xi_{\vec{p}}=0}^x=1/2$. These restrictions prevent the appearance of DQPTs. In the limit where $\Delta_i\to 0$ and the initial state  becomes the normal ground state  a discontinuity appears in the spin distribution allowing for the spins near the Fermi surface to be flipped and a DQPT to occur.
The rich dynamical behaviour of the order parameter is therefore not captured by any change in the presence of DQPTs. The transitions between the dynamical phases are continuous, a feature which is exhibited also by the Loschmidt echo.

Looking only at the initial state and the ground state of the final Hamiltonian, we see that there are no DQPTs for superconductor $\to$ superconductor quenches and a single DQPT for the normal $\to$ superconductor quench from the quantum critical point $g_i=0$. This seems to agree with the original DQPT proposal~\cite{HeylPolKeh}. On the other hand, we will find in what follows that DQPTs can also occur for quenches within the same equilibrium phase.

The time dependent Cooper pair distribution, $\gamma[\xi_{\vec{p}},t]$ can be calculated at large times throughout the phase diagram. As a particular case we can consider  Phase II. Expressions \re{asymptotespin} and \re{Aexp} hold in Phase II as well with the replacement of $\vec{\sigma}_{\vec{p}}$ with the unit vector along the magnetic field $B_{\vec{p}}$. This implies that $\cos{[\theta_{\vec{p}}]}$ in \eqref{Aexp} is  the long time asymptote of $\gamma[\xi_{\vec{p}},t]$ in Phase II    from which one can confirm that no zeros appear in the distribution function.   Furthermore, it is possible to check that under time translation of the initial state no DQPTs are generated. This is again due to the presence of the dephasing term which prevents the spin becoming flipped relative to its initial position.

\section{Soliton Dynamics}
\label{solitonsec}

 Here we  use DQPTs to examine the dynamics  which results when the initial state is taken to be an unstable stationary state of $\hat{H}(g_f)$~\cite{YuzbashTsyplyAltshuler,Yuzbash}. Whether a particular stationary state is stable or unstable is determined by linearizing the equations of motion about that solution. A stable stationary state is one in which the frequencies of the normal modes are only real. For example, linearizing~\eqref{uEOM} about the ground state of $\hat{H}(g_f)$ one finds that the solution has frequencies $\omega_{\vec{p}}=2\sqrt{\xi_{\vec{p}}^2+\Delta_f^2}$ coinciding with the spectrum of excitations of the BCS condensate [excited Cooper pairs]. It is natural to associate these to eigenstates of $\hat{H}(g_f)$ and the resulting dynamics are trivial. Unstable solutions on the other hand exhibit imaginary frequencies and therefore infinitesimal perturbations along these directions lead to exponential departure from the stationary state. Unstable stationary states can be either normal or anomalous.

If the initial state is normal, then this can be considered a quench from the free Fermi gas as we did in the previous section when the initial state was the normal ground state. For an anomalous initial state, since the initial and final values of the coupling are the same, $g_i=g_f$, i.e., such a situation is not an interaction quench of the type we considered above. Nevertheless it leads to nontrivial, far from equilibrium dynamics of $\Delta(t)$.  Such time evolution is a multi-soliton, meaning that it connects unstable stationary states at $t=-\infty$ and $t=\infty$ and furthermore  can be decomposed into sums of single soliton solutions in a certain limit. Dynamics of this type are a feature of the interacting model and are completely absent from the non-interacting system.

Interestingly, we find that the number of DQPTs is related to the soliton number. Normal $k$-soliton solutions belong to the class
of states discussed in Sect.~\ref{1normalsoliton} -- states where the imaginary part of the anomalous Green's function vanishes on the Fermi surface. Their Cooper pair distribution  $\gamma[\xi_{\vec{p}}, t]$ has a permanent  zero at the Fermi energy and the angle of rotation of spins 
  near the Fermi surface is $\Phi(\infty, -\infty)=2\pi k$, which means that there are exactly $k$ DQPTs. Anomalous single solitons are of two types, $\Delta_+$ and $\Delta_-$ [see below], and the total number of solitons in a multi-soliton solution is $k=k_++k_-$, where $k_+(k_-)$ is the number of $\Delta_+(\Delta_-)$ solitons. In this case zeros of  $\gamma[\xi_{\vec{p}}, t]$ are transient, their positions are time dependent, and there are $k_-$ DQPTs.

We note also that DQPTs for single normal and anomalous solitons have several interesting properties not shared with more general solutions. These properties are:
\begin{enumerate}

\item There is a single DQPT that occurs at the global maximum  of $|\Delta(t)|$ and  $|\Delta(t_\mathrm{DQPT})|=\Delta_f$.  
\label{property1}

\item At the DQPT point $s^y_{\vec{p}}=0$ for all $\vec{p}$ and all  $s^x_{\vec{p}}$ have the same sign which is known as phase locking.
\label{locking}

\item One consequence of property~\ref{locking} is that all separation variables are real at $t=t_\mathrm{DQPT}$ as one can show using
\eqref{sepv}.
\label{cor1}

\item Another consequence of property~\ref{locking} is that the Bogoliubov amplitudes $u_{\vec{p}}$ and $v_{\vec{p}}$ are real at $t=t_\mathrm{DQPT}$. This implies that the state of the system~\re{psit} is time-reversal invariant with respect to the DQPT point,
$\Psi_\mathrm{BCS}^*( t)=\Psi_\mathrm{BCS}(2 t_\mathrm{DQPT}-t)$.
\label{cor2}

\end{enumerate}

\noindent Property~\ref{property1} in the case of the normal soliton we established in Sec.~\ref{1normalsoliton}, while property~\ref{locking} follows 
from~\eqref{spin1soliton} and $\dot \Delta(t_\mathrm{DQPT})=0$. For the single anomalous soliton we prove these properties later in this section.

\subsection{Normal Solitons}

The simplest unstable stationary state of $H(g_f)$ is the  Fermi gas ground state. We have already seen that in this case the order parameter exhibits a single soliton peak with an accompanying DQPT, which occurs due to the flipping of spins close to the Fermi surface where the initial spin distribution had a discontinuity.  

Excited states of the Fermi gas are also unstable stationary states and lead to multi-soliton dynamics. For all these states 
$s^x_{\vec{p}}=0$ at the Fermi surface and therefore the Cooper distribution  has a permanent zero at the Fermi energy. Similar to the  Fermi gas ground state this leads to DQPTs. The total number of DQPTs depends on the total angle of rotation  
$\Phi_\mathrm{tot}=2\int_{-\infty}^\infty dt\Delta(t)$ around the $x$-axis of spins $\vec{s}_\pm$ near the Fermi surface, see~\eqref{Phi}. 
For the single normal solitons we saw that $\Phi_\mathrm{tot}=2\pi$. Next we investigate DQPTs for normal multi-solitons.

 The number of solitons in $\Delta(t)$ which emerge is related to the number of discontinuities in the spin distribution. For $2k-1$ discontinuities in the spin distribution, the spectral polynomial $Q_{2n}(u)$ has up to $k$ complex conjugate pairs of roots and the dynamics of $\Delta(t)$ consists of up to $k$ solitons. In general the multi-soliton solutions can result in a complicated $\Delta(t)$ with interference fringes from overlapping solitons, see  Fig.~\ref{fig:two_norm_sol_a_gt_D4_far_D}. Despite this however the Loschmidt echo displays exactly $k$ DQPTs each of which is of the same form as the single soliton case. 
 
 \begin{figure}[hbt]
    \centering%
    \subfloat[  ] 
    {\includegraphics[width=.9\columnwidth]{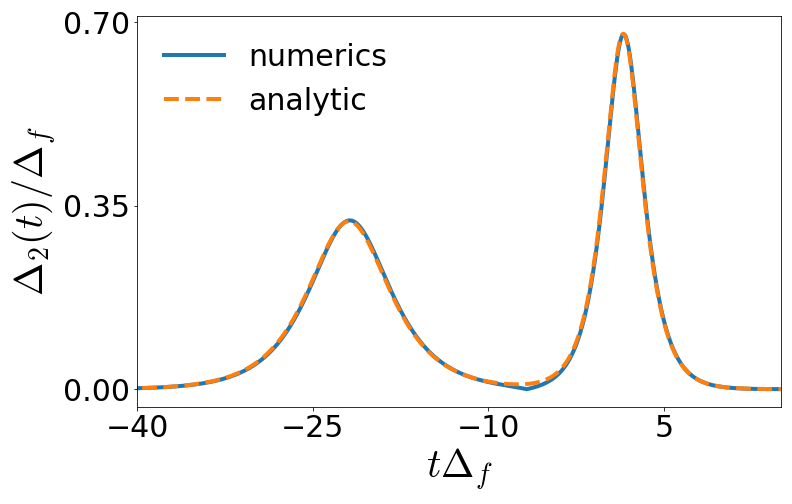}}\\
    \subfloat[] 
    {\includegraphics[width=.9\columnwidth]{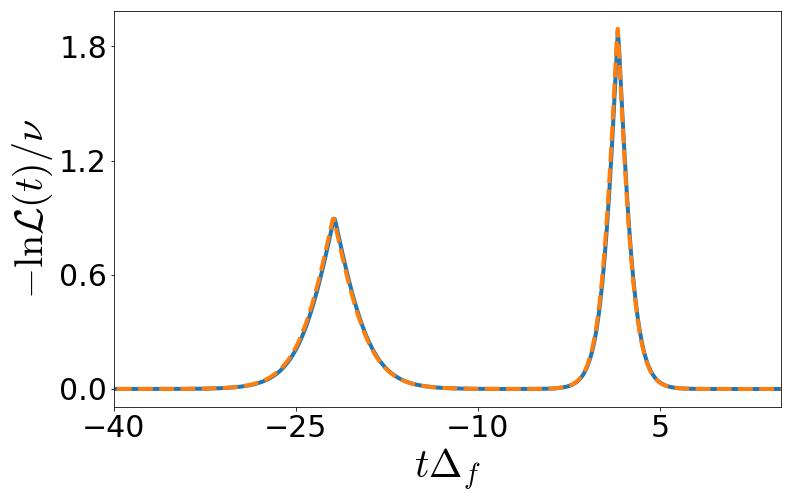}}
    \caption{(a) Order parameter $\Delta(t)$ and (b)  log of the Loschmidt echo for a normal 2-soliton with $a < \Delta_f/4$ in the regime where  the 2-soliton solution reduces to a simple sum of two single solitons. As a result, the echo shows two DQPTs and  its log is a sum of logs of two single soliton Loschmidt echos. Here $\Delta(t)$ and $\ln{\cal L}(t)$ are evaluated from~Eqs.~\re{Delta2soliton} and \re{2normalsolitonecho} [`analytic'], respectively, and compared with direct numerical simulation [`numerics'] of spin equations of motion starting from initial conditions~\re{2solbc}.  }
    \label{fig:two_norm_sol_a_lt_D4_far_F}
\end{figure}

\begin{figure}[hbt]
    \centering%
    \subfloat[  \label{fig:two_norm_sol_a_gt_D4_far_D}]{\includegraphics[width=.9\columnwidth]{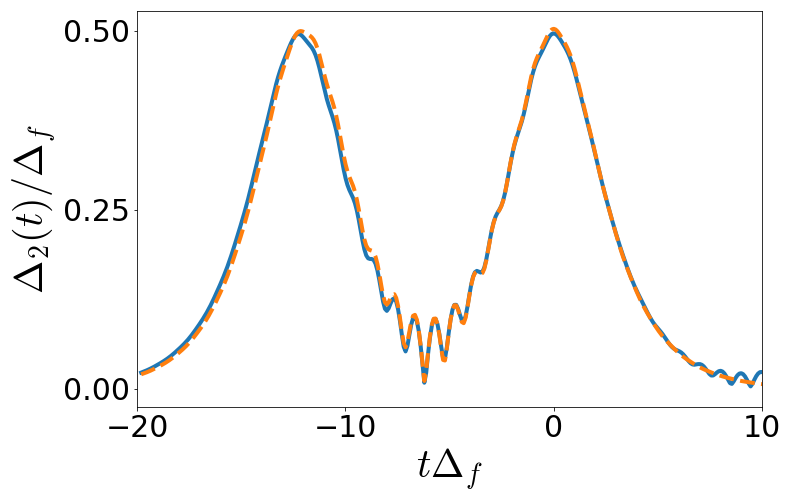}}\\%
    \subfloat[ \label{fig:two_norm_sol_a_gt_D4_far_F}]{\includegraphics[width=.9\columnwidth]{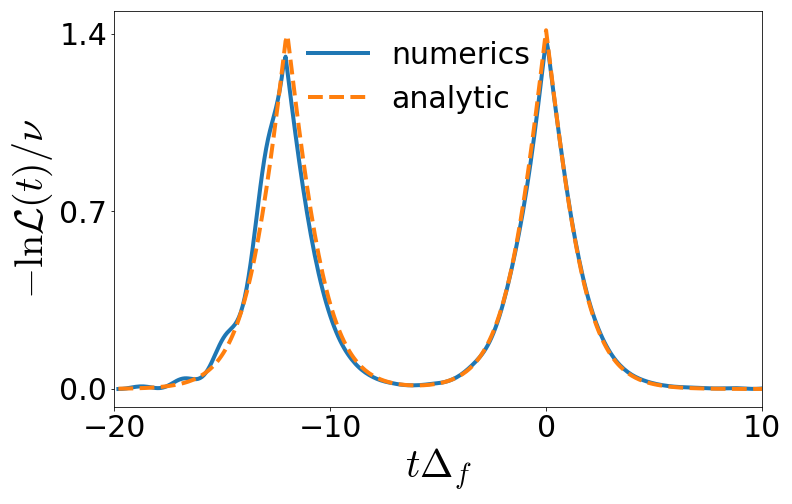}}
    \caption{(a) Order parameter $\Delta(t)$ and (b)  log of the  Loschmidt echo for a normal 2-soliton with $a > \Delta_f/4$. All parameters and the meaning of `analytic' vs. `numerics' are the same as in Fig.~\ref{fig:two_norm_sol_a_lt_D4_far_F}, except  $a=1.5$,
    $\tau_2=-13.3$, $t_0=-22$, and the initial conditions for simulations are in \eqref{2solbca}.  In the time interval where the two solitons  overlap, the phase difference between them leads to complicated interference fringes with many local minima and maxima. Nevertheless, the Loschmidt echo shows two DQPTs, one associated with each soliton.}
    \label{fig:two_normal_soliton_a_gt_D4}
\end{figure}

To examine this explicitly we investigate the dynamics from the excited state described by the choice $e_{\vec{p}}=-\text{sgn}[\xi_{\vec{p}}(\xi_{\vec{p}}^2-a^2)]$. Evaluating $\vec{L}^2(u)$ for this initial state we find that $Q_{2n}(u)$ has $n-4$ real  and two complex conjugate pairs of double roots~\cite{Yuzbash}. The complex pairs are denoted $\pm i\eta_{1,2}$ with
\begin{eqnarray}
\eta_{1,2}=\frac{\Delta_f}{4}\pm\sqrt{\frac{\Delta^2_f}{16}-a^2}.
\end{eqnarray}
 where again $\Delta_f=2De^{-1/V\nu g_f}$. The equations of motion reduce to a system of three variables $u_{j}(t)$, $j=1,2,3$ and  subsequently $\Delta(t)$ and $\vec{s}_{\vec{p}}(t)$ can be determined. We note that the roots are purely imaginary for $a<\Delta_f/4$ and otherwise have a real part with either case needing to be separately considered. For $a<\Delta_f/4$ it is found that the order parameter is~\cite{Yuzbash}
 \begin{equation}
 \label{Delta2soliton}
 \begin{split}
 \Delta_2(t)=A\left|\frac{h(t)}{h(t)\ddot{h}(t)-\dot{h}^2(t)}\right|,\\
 h(t)=\sum_{j=1}^2\frac{\cosh{(2\eta_j (t-\tau_j))}}{2\eta_j}e^{i\phi_j}.
 \end{split}
 \end{equation}
Here $\tau_{1,2}$ and $\phi_{1,2}$ are constants which depend on how the unstable state was perturbed and $A=4|\eta_1^2-\eta_2^2|$. Specifically, the deviation from the unstable normal state that produces this solution is
\beg
\begin{split}
s_\p^-=\frac{2e_\p(\eta_1+\eta_2)}{\eta_1-\eta_2}\left[ \frac{\eta_1 e^{2\eta_1(t_0-\tau_1)} }{ \eta_1-i\xi_\p} - \frac{\eta_2 e^{2\eta_2(t_0-\tau_2)} }{ \eta_2-i\xi_\p}\right],\\
s_\p^z=\frac{e_\p}{2} \sqrt{1-4|s_\p^-|^2},\quad e_{\vec{p}}=-\text{sgn}[\xi_{\vec{p}}(\xi_{\vec{p}}^2-a^2)],
\end{split}
\label{2solbc}
\en
where $\Delta_f t_0\gg1$ and we set $\phi_{1,2}=0$.

The 2-soliton nature of this solution is manifest when one takes $|\tau_1-\tau_2|\gg \Delta_f^{-1}$. In this limit we have that 
\begin{equation}
\begin{split}
\Delta_{2}(t)\approx\Delta_{1,1}(t)+\Delta_{1,2}(t),\\
\Delta_{1,j}(t)=\frac{2\eta_j}{\cosh{(2\eta_j(t-\tau'_j))}}.
\end{split}
\label{2prime}
\end{equation}
where $\tau'_1=\tau_1+\tau_0$, $\tau'_2=\tau_2-\tau_0$  with $\tau_0=\mathrm{sgn}(\tau_1-\tau_2)\tanh^{-1}[2\eta_1\eta_2(\eta_1^2+\eta_2^2)^{-1}]$ being a constant shift.  The 2-soliton solution therefore reduces to a sum of two widely separated single soliton solutions. 

The spin distribution also separates in this manner and the normal 2-soliton Loschmidt echo $\mathcal{L}_\mathrm{2s}(t)$ in this limit is found to be
\begin{equation}
\begin{split}
\ln \mathcal{L}_\mathrm{2s}(t)= \ln\mathcal{L}_\mathrm{1s}(t, \eta_1, \tau'_1)+\ln \mathcal{L}_\mathrm{1s}(t, \eta_2, \tau'_2),
\end{split}
\label{2normalsolitonecho}
\end{equation}
where $\mathcal{L}_\mathrm{1s}$ is the single soliton echo given by \eqref{singleEcho}. We double check this answer in  Fig.~\ref{fig:two_norm_sol_a_lt_D4_far_F} by comparing  it to a direct numerical 
 run of equations of motion using initial conditions~\re{2solbc},  \eqref{SpinEcho}, and  parameters $\Delta_f=0.9, a=0.21$,   $\tau_1=0, \tau_2=-20.8$, $t_0=-50$, and $n=2\times 10^5$ spins. 
 
 Therefore, each soliton is accompanied by a DQPT which, like in the single soliton case,  coincides with the peak of each soliton in this limit.  Now the spins  near the Fermi surface
complete two rotations by $2\pi$, one for each soliton, i.e., $\Phi_\mathrm{tot}=4\pi$. Note however that  property~\ref{locking} and
its consequences hold only approximately. This is clear from the fact that the 2-soliton is not time-reversal symmetric with respect to the maximum of either soliton. This symmetry emerges only when  $|\tau_1-\tau_2|\to\infty$ sending one of the solitons to infinity.

When $a>\Delta_f/4$ the complex roots are given by $\pm \mu\pm i\eta$ with $\mu=\sqrt{a^2-\Delta_f^2/16}$ and $\eta=\Delta_f/4$. With this initial state the order parameter is also given by \eqref{Delta2soliton} but with $A=16\mu \sqrt{\mu^2+\eta^2}$ and
\begin{equation}
\begin{split}
h(t)=e^{-2i\mu t +i\phi_1}\frac{\cosh{2\eta(t-\tau_1-i\beta)}}{2\eta}\\
+e^{2i\mu t +i\phi_2}\frac{\cosh{2\eta(t-\tau_2+i\beta)}}{2\eta}
\end{split}
\end{equation}
where $\beta=\arctan{(\mu/\eta)}$. 
The initial deviation from the unstable state for this 2-soliton solution is
\beg
s_\p^-=-\frac{i e_\p\eta e^{2 \gamma t_0} }{\mu} \left[\frac{(\mu+i\eta)e^{-2 i\mu t_0}}{\xi_\p +\mu+i\eta} +\frac{(\mu-i\eta)e^{2 i\mu t_0} }{\xi_\p -\mu+i\eta}\right],
\label{2solbca}
\en
with $s_\p^z=\frac{e_\p}{2} \sqrt{1-4|s_\p^-|^2}$ as before.

Here again one can take the limit $|\tau_1-\tau_2|\gg \Delta_f^{-1}$ and find that $\Delta_2(t)$ becomes the sum of two single solitons. The added feature is that these solitons rotate with respect to each other with frequency $4\mu$. This phase difference between the two leads to a complicated profile of interference fringes when the two solitons are close to each other, see  Fig.~\ref{fig:two_norm_sol_a_gt_D4_far_D}. Despite this there are still exactly two DQPTs one associated with each soliton as shown in  Fig.~\ref{fig:two_norm_sol_a_gt_D4_far_F}. Decreasing the separation between the solitons, we always observe 2 DQPTs. Only when
$\tau_1-\tau_2=0$ do they merge into a single singulatiry, which can be probably considered a doubly degenerate DQPT.  In the widely separated limit the Loschmidt echo is still given by \eqref{2normalsolitonecho} and the order parameter by \eqref{2prime} with $\eta_1\to\eta$ and $\eta_2\to\eta$.

The same story persists for higher soliton number. For $2k-1$ discontinuities in the spin distribution, the order parameter exhibits
up to $k$ solitons and associated with each of these is a single DQPT. In the limit where these solitons are widely separated the DQPTs occur exactly when they are at their peak. When closer together complicated interference patterns may appear in the profile of $|\Delta(t)|$ and the DQPTs no longer coincide with peaks of the order parameter. Their number however remains equal to the soliton number and the total angle of rotation of spins on either side of the discontinuity at the Fermi level is $\Phi_\mathrm{tot}=2\pi k$. In all cases the DQPTs are transient and can be removed by time translation of the initial state and  are due to the zero in the Cooper pair distribution  at the Fermi energy.



\subsection{Anomalous Solitons}

There also exist unstable anomalous states corresponding to excited states of a superconductor. Such states are described by $2k$ discontinuities in their spin distribution and result in anomalous $k$-solitons. Here the dynamics are more subtle and solitons which emerge from such states come in two types denoted $\pm$ which either are ($-$) or are not ($+$) accompanied by a single DQPT.
   Similar to the superconducting ground state, the spin at the Fermi energy is static for these initial states. Indeed, we see  from~\eqref{AnomSpin} that $\vec{s}_{ \xi_\p=0}=\pm \hat x/2\ne 0$ and the spin texture is continous at the Fermi surface. There is no interaction change involved, so $g_i=g_f$ and $\Delta_i=\Delta_f$; we will use $\Delta_f$ to denote the ground state gap.

We shall consider in detail only the simplest anomalous unstable state which is described by \eqref{AnomSpin} with the choice $e_{\vec{p}}=\text{sgn}(|\xi_{\vec{p}}|-a)$. 
Inserting this distribution into the self consistency condition, $\Delta_\mathrm{in}=g_i\sum_{\vec{p}} s^x_{\vec{p}}(0)$ one finds that the solutions for the initial value of the order parameter, $\Delta_\mathrm{in}$, are governed by
\begin{eqnarray}
\Delta_\mathrm{in}(\Delta_\mathrm{in}-\Delta_f)^2=4a^2\Delta_f. 
\end{eqnarray}
 Provided $3\sqrt{3}a<\Delta_f$ there are two physical solutions which have $0\le \Delta_\mathrm{in}<\Delta_f$. The larger of the two solutions  obeys $\Delta_f/3<\Delta_\mathrm{in}<\Delta_f$  and so is continuously connected to the ground state solution, $\Delta_f$, by reducing $a$. These solutions can be identified with the stable stationary states and therefore the true quantum eigenstates of $\hat{H}(g_f)$. The smaller solutions, $\Delta_\mathrm{in}<\Delta_f/3$, represent the unstable states. They are continuously connected to the normal ground state, discussed above, by reducing $a$ and satisfy the relation
\begin{eqnarray}
a=\frac{1}{2}\sqrt{\frac{\Delta_\mathrm{in}}{\Delta_f}}(\Delta_f-\Delta_\mathrm{in}).
\end{eqnarray}
Using the unstable state one finds that $Q_{2n}(u)$ has $n-3$ doubly degenerate real roots, a pair of complex roots at $\pm i \Delta_\mathrm{in}$ and a pair of doubly degenerate roots at $\pm i (\Delta_f-\Delta_\mathrm{in})/2$. The dynamics of the system reduces to two variables $u_j(t)$, $j=1,2$ governed by \eqref{uEOM}.  From this the order parameter and spin distribution can be determined but in contrast to normal anomalous state there are two types of soliton which can occur. They are described by~\cite{Yuzbash}
\begin{equation}
\begin{split}
\Delta_\pm(t)-\Delta_\mathrm{in}&=\frac{\lambda^2}{2\Delta_\mathrm{in}\pm (\Delta_f-\Delta_\mathrm{in})\cosh{[\lambda (t-\tau)]}},\\
&\lambda=\sqrt{(\Delta_f-\Delta_\mathrm{in})^2-4\Delta_\mathrm{in}^2}.
\end{split}
\end{equation}
Both $\Delta_+$ and $\Delta_-$ solutions reproduce the normal soliton in the limit $\Delta_\mathrm{in}\to 0$. Away from this limit, there are two notable distinctions between these two solitons. First, $\Delta_+(t)$ is always positive, while $\Delta_-(t)$ changes sign twice at 
$t=\tau\pm t_*$, where
\begin{equation}
t_*=\cosh^{-1} \frac{ (\Delta_f-\Delta_\mathrm{in})^2-2\Delta_\mathrm{in}^2}{\Delta_\mathrm{in}(\Delta_f-\Delta_\mathrm{in})},
\label{tstar}
\end{equation}
as it evolves from $\Delta_-(-\infty)=\Delta_\mathrm{in}$ to $\Delta_-(\tau)=-\Delta_f$ and back to $\Delta_-(+\infty)=\Delta_\mathrm{in}$. Second, similar to the single normal soliton $|\Delta_-(t)|$  reaches the ground state value $\Delta_f$  at its peak whereas $\Delta_+(\tau)=\Delta_f-\Delta_\mathrm{in}$. 

The time dependent spin configuration for the anomalous solitons is 
\beg
\begin{array}{l}
\dis s_{\vec{p}}^x(t)=\frac{e_{\vec{p}}\xi_{\vec{p}}^2(\Delta_\pm(t)-\Delta_\mathrm{in})}{2(\xi_{\vec{p}}^2+\gamma^2)\sqrt{\xi_{\vec{p}}^2+\Delta_\mathrm{in}^2}}+
\frac{e_{\vec{p}}\Delta_\mathrm{in}}{2\sqrt{\xi_{\vec{p}}^2+\Delta_\mathrm{in}^2}},\\
\\
\dis s_{\vec{p}}^y(t)=-\frac{e_{\vec{p}}\xi_{\vec{p}} \dot\Delta_\pm(t)}{4(\xi_{\vec{p}}^2+\gamma^2)\sqrt{\xi_{\vec{p}}^2+\Delta_\mathrm{in}^2}},\\
\\
\dis s_{\vec{p}}^z(t)=\frac{e_{\vec{p}}\xi_{\vec{p}}(\Delta_\pm^2(t)-\Delta_\mathrm{in}^2)}{4(\xi_{\vec{p}}^2+\gamma^2)\sqrt{\xi_{\vec{p}}^2+\Delta_\mathrm{in}^2}}-
\frac{e_{\vec{p}}\xi_{\vec{p}}}{2\sqrt{\xi_{\vec{p}}^2+\Delta_\mathrm{in}^2}},\\
\end{array}
\label{indvsp}
\en
where $2\gamma=\Delta_f-\Delta_\mathrm{in}$.  

\begin{figure}[hbt]
    \centering
    \subfloat[$\Delta_\mathrm{in} = 6.48\cdot10^{-3}$\label{fig:one_anom_sol_Dplus_D}]{\includegraphics[width=.5\columnwidth]{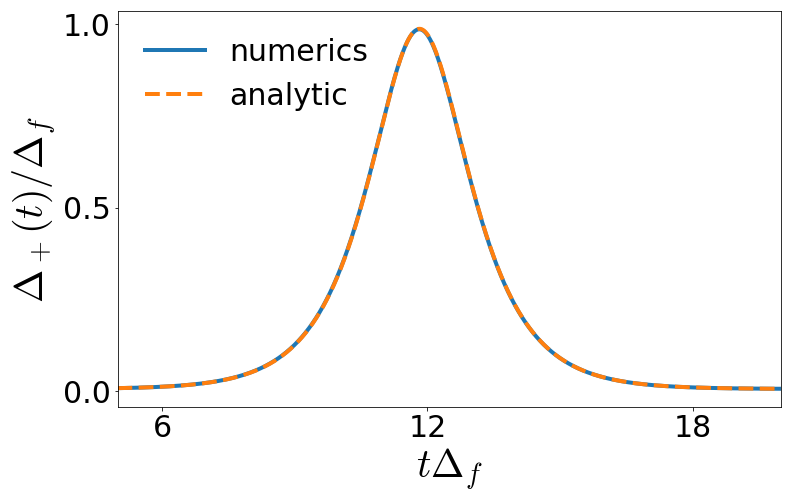}}
    \subfloat[$\Delta_\mathrm{in} = 6.48\cdot10^{-3}$\label{fig:one_anom_sol_Dplus_F}]{\includegraphics[width=.5\columnwidth]{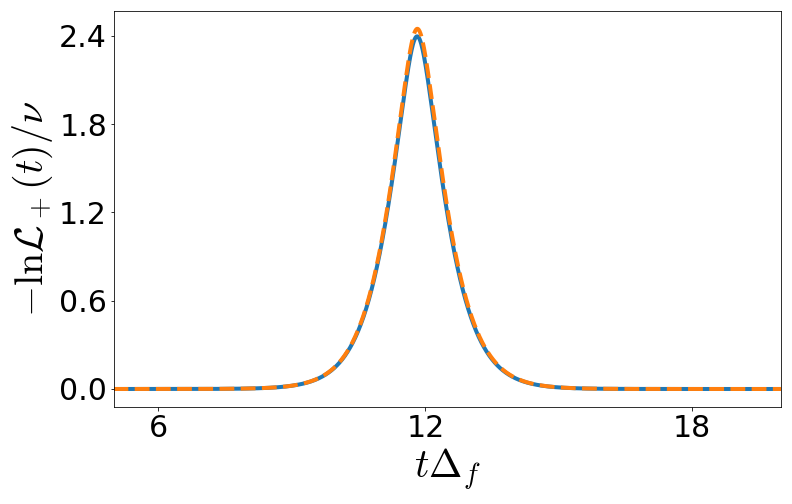}} \\
    \subfloat[$\Delta_\mathrm{in} = 3.63\cdot10^{-3}$\label{fig:one_anom_sol_Dminus_D}]{\includegraphics[width=.5\columnwidth]{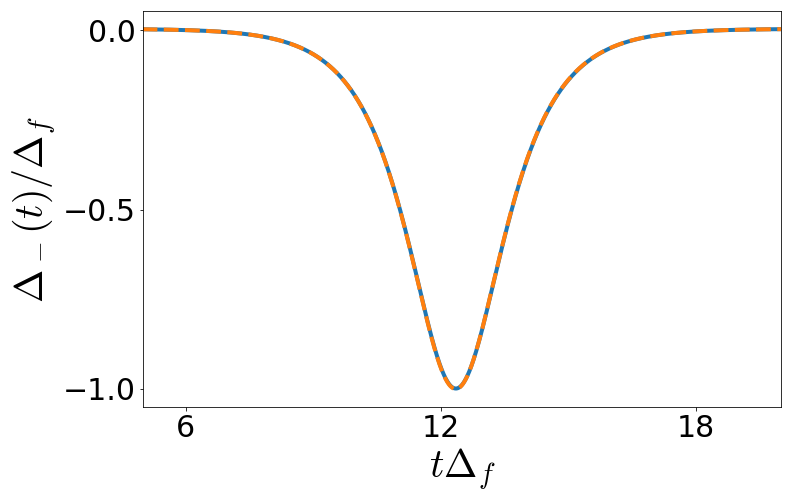}}
    \subfloat[$\Delta_\mathrm{in} = 3.63\cdot10^{-3}$\label{fig:one_anom_sol_Dminus_F}]{\includegraphics[width=.5\columnwidth]{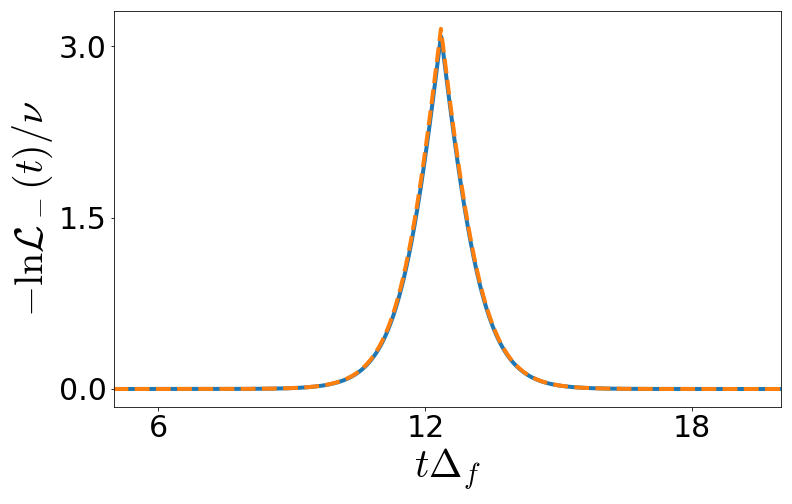}}
    \caption{Order parameter and Loschmidt echo for the two varieties of a single anomalous soliton. Numerical simulation with $n=2\times 10^5$  spins and $\Delta_f=1$ are compared to the analytic result~\re{AnomEcho}.  For (a) $\Delta_+$ soliton there are no DQPTs  in its Loschmidt echo as shown in (b). For (c) $\Delta_-$ soliton there is a DQPT  when 
    $|\Delta_-(t)|$ reaches the ground state value  $\Delta_f$ as shown in (d). }
    \label{fig:one_anom_sol}
\end{figure}

  The Loschmidt echo can now be evaluated with the result 
\begin{eqnarray}\label{AnomEcho}
\mathcal{L}_\pm(t)=e^{-\pi\nu V(\Delta_f+\Delta_\mathrm{in})\left[1-\sqrt{1-\left(\frac{\Delta_\pm(t)-\Delta_\mathrm{in}}{\Delta_f+\Delta_\mathrm{in}}\right)^2}\right]}.
\end{eqnarray}
Again a DQPT occurs when the argument of the square root in \eqref{AnomEcho} vanishes however this is possible only for $\Delta_-(t)$ and not $\Delta_+(t)$. Therefore in the anomalous case DQPTs are associated only with one of the two possible solutions. In particular a DQPT occurs when the order parameter magnitude hits the ground state value $\Delta_-(\tau)=-\Delta_f$. Expanding about this point we see that the DQPT is of the same form as in the normal soliton given in \eqref{logEcho}. Plots of  $\Delta_\pm(t)$ and the Loschmidt echo are shown in  Fig.~\ref{fig:one_anom_sol}.

Let us also show that property~\ref{locking} stated above (phase locking) holds for $\Delta_-$ and not for $\Delta_+$. Since $\dot \Delta_\pm(\tau)=0$,
$s_{\vec{p}}^y(\tau)\equiv0$ for both solitons. However, $s_{\vec{p}}^x(\tau)$ is discontinuous and changes sign at $\xi_{\vec{p}}=\pm a$
in the case of $\Delta_+$, while for $\Delta_-$ we find from~\eqref{indvsp}
\beg
\dis s_{\vec{p}}^x(\tau)=\frac{  \Delta_f \text{sgn}(|\xi_{\vec{p}}|-a) ( a^2- \xi_{\vec{p}}^2 )}{2(\xi_{\vec{p}}^2+\gamma^2)\sqrt{\xi_{\vec{p}}^2+\Delta_\mathrm{in}^2}}\le 0.
\en
Discontinuities at $\xi_\p=\pm a$ otherwise permanently present for both solitons disappear at the DQPT  due to the vanishing
of $s^x(\xi_\p=\pm a)$. Remarkably, this happens only at the DQPT point and only for the $\Delta_-$ soliton. 
As mentioned in the beginning of this section, the fact that all $s_{\vec{p}}^y(\tau)$ vanish and all $s_{\vec{p}}^x(\tau)$ are of the same sign also means that the DQPT for $\Delta_-(t)$ coincides with the variables   $u_{1,2}(t)$ both crossing the real axis. On the other hand, for $\Delta_+(t)$,  $u_{1,2}(t)$ do not become real and no DQPT occurs. 
  
  In addition one can compute the Cooper pair distribution~\re{CPDF}   in both cases and find that  zeros only appear for $\Delta_-(t)$ and not $\Delta_+(t)$.    Since   $\gamma[\xi_{\vec{p}}, t]$ is an even function of $\xi_{\vec{p}}$  due to particle-hole symmetry, for any zero at $\xi_0\ne 0$, there is also a zero at $-\xi_0$.  
The condition $\gamma[\xi_0,t]=0$ reads
 \beg
 s^x(\xi_0, t)\Delta(t)=s^z(\xi_0, t) \xi_0.
 \label{CPDc}
 \en
 For single anomalous solitons $\Delta_\pm(t)$ this reduces to a biquadratic equation
  \beg
  \begin{split}
  4\xi_0^4+\xi_0^2 \left( [\Delta_\pm(t)+\Delta_\mathrm{in}]^2+ (\Delta_f-\Delta_\mathrm{in})^2\right)+\\
   +(\Delta_f-\Delta_\mathrm{in})^2\Delta_\mathrm{in}\Delta_\pm(t)=0.
  \end{split}
  \en 
  This equation has real solutions only when $\Delta_\pm(t)\le 0$, which is only possible for $\Delta_-(t)$. There is a pair of zeros when $\Delta_-(t)\le 0$ that emerge and disappear together when $\Delta_-(t)$ crosses the real axis.  They
  first emerge as a doubly degenerate zero at the Fermi level,  $\xi_0=0$, when   $\Delta_-(t)$  vanishes for the first time at $t=\tau-t_*$, where $t_*$ is given by \eqref{tstar}. The two zeros then symmetrically move away from the Fermi level in opposite directions until they reach their extremal positions $\pm\xi_{\max}$ with $2\xi_{\max}^2=(\Delta_f-\Delta_\mathrm{in})\Delta_\mathrm{in}$ at the DQPT time $t=\tau$. After this the two zeros turn
  around, return to the Fermi level at $t=\tau+t_*$, and disappear altogether for $t> \tau+t_*$. The Fermi level thus acts as a source and
  sink for the zeros of the Cooper pair distribution.
  This example shows that not only are DQPTs associated with the presence of zeros in $\gamma[\xi_{\vec{p}},t]$, but also that zeros away from the Fermi level are not protected by the particle-hole symmetry and can appear, disappear, and move around in time.

 The anomalous $k$-solitons can also be constructed and when they are widely separated simplify to a sum of single anomalous solitons which may be of either $\Delta_\pm(t)$ type. For any $\Delta_-(t)$  type soliton present in the sum there will be a corresponding DQPT which occurs when the soliton hits its peak. As with the normal $k$-solitons the number of DQPTs does not change when the solitons are not widely spaced.
 
Once again we have seen that the presence of interactions, encoded by the self consistency condition, allows for DQPTs which would not be expected from the non-interacting analysis. As with the previous section the number of DQPTs depends upon the number of discontinuities in the initial spin distribution, occurs in conjunction with zeros appearing in the Cooper pair distribution  and in a further departure from the noninteracting system can occur when the initial state is anomalous. 

\subsection{Soliton Train}

There also exist other solutions to the equations of motion which do not emerge from stationary states of the Hamiltonian. In general, the solution to the self consistent equations of motion \re{uEOM} can be written as  hyperelliptic functions with $n$ incommensurate basic frequencies~\cite{YuzbashAltKuzEnol}. These are related to the solitons discussed above by taking the limit where all discrete frequencies vanish and as the result the hyperelliptic functions reduce to elementary functions.  They can be thought of as nonequilibrium steady state solutions of \eqref{spinEOM} in which the order parameter exhibits persistent multi-periodic oscillations. 

The simplest of these was discovered in \cite{BarakovSpivakLevitov}. It is given by 
\beg
\begin{split}
&s^z_{\vec{p}}(t)=-\frac{\text{sgn}[\xi_{\vec{p}}]}{2} \sigma_{\vec{p}}^z(t),\quad  s^-_{\vec{p}}(t)=-\frac{\text{sgn}[\xi_{\vec{p}}]}{2} \sigma_{\vec{p}}^-(t),\\
 &\Delta(t)=\delta_+\text{dn}\,[\delta_+(t-\tau),k],
 \end{split}
 \label{train}
 \en
 with  $\vec{\sigma}_{\vec{p}}$ given by~\eqref{sigmazm}. We immediately see from~\eqref{sigmazm} that $s^x_{\vec{p}}=0$ at the Fermi surface. As discussed above, this implies that the Cooper pair distribution  $\gamma[\xi_{\vec{p}}, t]$ has a permanent zero at $\xi_{\vec{p}}=0$, which causes DQPTs.
 In this case, since $\Delta(t)$ is periodic, the angle of rotation~\re{Phi} of spins near the Fermi surface is unbounded and therefore
 there are infinitely many DQPTs in the limit $t\to\infty$. Specifically, the angle of rotation is
 \beg
 \Phi(t,t_i)=\frac{2\pi}{T}(t-t_i)+F(t)-F(t_i),
 \en
 where $T$ is the period of $\Delta(t)$ and $F(t)$ is a periodic function   with the same period. To derive this equation, we
 used the Fourier series of the Jacobi elliptic function dn. A DQPT occurs each time
 $\Phi(t,t_i)$ equals an odd multiple of $\pi$, i.e., at $t_\mathrm{DQPT}=t_i+(m+\frac{1}{2})T$ with integer $m$, because this means  spins in the vicinity of the Fermi surface are inverted as compared
 to $t=t_i$. We see that DQPTs occur periodically with period $T$, same as the period of  $\Delta(t)$.
 
 To evaluate the Loschmidt echo for this solution we take the initial state to be given by \eqref{sigmazm}  at the point where $\Delta(t)=\delta_-$, which corresponds to $t_i=\tau+T/2$. The  echo can then be calculated to be
\begin{eqnarray}\nonumber
\mathcal{L}(t)&=&e^{-\pi \delta_+ \nu V\left[2-\sqrt{a(t)+2\sqrt{b(t)}}\right]},
\end{eqnarray}
where $a(t)=\left(1+{\delta_-}/{\delta_+}\right)^2+1-\left({\Delta(t)}/{\delta_+}\right)^2$ and $b(t)= \left(1+{\delta_-}/{\delta_+}\right)^2(1-\left[{\Delta(t)}/{\delta_+}\right]^2)$. This shows that a DQPT occurs each time the order parameter is at a maximum, $\Delta(t)=\delta_+$, which is at $t_\mathrm{DQPT}=\tau+mT$. Each DQPT is of the form of that for the single soliton and can be traced back to the fact the the spins close to the Fermi surface become flipped when $\Delta(t)$ goes from a minimum to a maximum.

 Similar to the normal initial states,  DQPTs here are  a consequence
of  the maximum discontinuity at the Fermi surface even though $\Delta(t)$ never vanishes for this solution.   Unlike the single normal soliton however, the phase locking property~\ref{locking} and its two corollaries    hold only when we choose an initial state that corresponds to the maximum or minimum of $\Delta(t)$. This once more underscores the fact that these properties are not properties 
of DQPTs per se, but rather   of  certain soliton-like peaks, which sometimes coincide with  DQPTs due to their enhanced symmetry.

In contrast to solitons, the DQPT here appears periodically and cannot be removed by translating the initial state forward in time.  Such purely periodic solutions however are finely tuned and are destroyed by perturbations of the initial state or the Hamiltonian. In addition,   making the imaginary part of the equal time anomalous Green's function ($s_\p^x$) nonzero on the Fermi surface will generally destroy  such DQPTs.  

Interestingly, DQPTs here serve as yet another marker of  the drastic difference between this unstable, finely tuned  soliton train solution and Phase III steady state considered in section~\ref{scinst}. Even though both solutions share the same functional form of $\Delta(t)$, there are no DQPTs in Phase III, which is a robust multi-periodic solution of the equations of motion and, similar to the BCS ground state, its spin distribution is a continuous function of energy $\xi_\p$.


\section{Test  initial states}

Thus far we have examined the dynamics of the system using some natural initial states, the ground and excited states of the BCS superconductor and from this a number of common features of DQPTs are evident. They occur when the initial state  exhibits a number of discontinuities in  its spin distribution and coincide with a zero appearing in the Cooper pair distribution. With the exception 
of anomalous solitons, all states that produced DQPTs were states with the  maximum possible discontinuity in the spin texture at the
Fermi surface. In the presence of particle-hole symmetry this feature alone, as we have proved, automatically leads to DQPTs.  
 One can then ask whether DQPTs can always be associated to discontinuities in the initial state and also whether or not all DQPTS are accompanied by zeros of the Cooper pair distribution. To investigate this further we have studied the dynamics for  an array of initial states specially designed to answer these questions.

\begin{figure}[hbt]
    \centering
    \subfloat[ \label{fig:sin_cos_D}]{\includegraphics[width=.5\columnwidth]{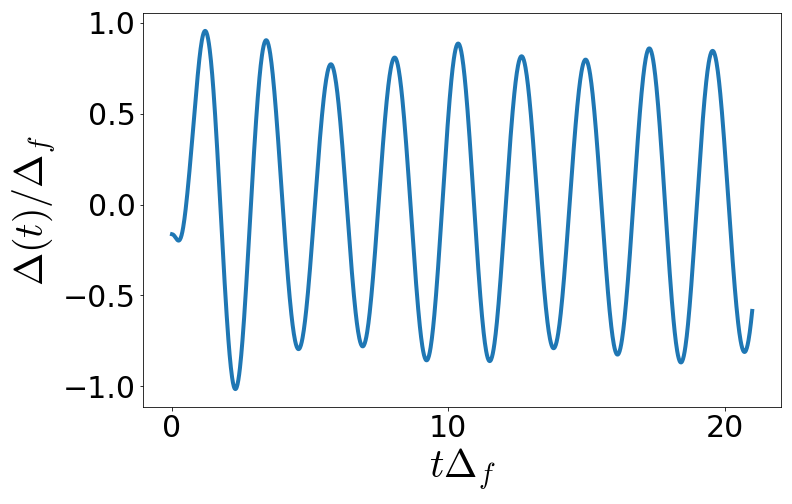}}
    \subfloat[ \label{fig:sin_cos_F}]{\includegraphics[width=.5\columnwidth]{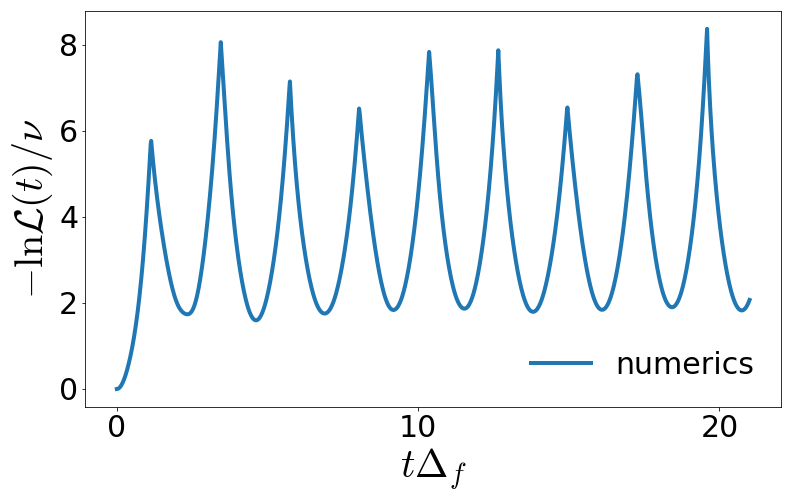}}
    \caption{(a) Order parameter $\Delta(t)$ and (b) log of the Loschmidt echo for the time evolution with the $s$-wave BCS Hamiltonian   starting from the `sin' initial state~\re{eq:sin_cos} . In contrast to solitons, here the pseudospin distribution  $\vec{s}_\p(t)$ is continuous in single-particle energy $\xi_\p$ at all times. Nevertheless, $s_\p^z(0)$  changes sign several times as a function of $\xi_\p$. This produces zeros in the Cooper pair distribution [see Fig.~\ref{fig:gammas}] and hence    DQPTs. }
     \label{sinfig}
\end{figure}

 \begin{figure}[hbt]
    \centering
    \subfloat[ \label{fig:three_tanh_D}]{\includegraphics[width=.5\columnwidth]{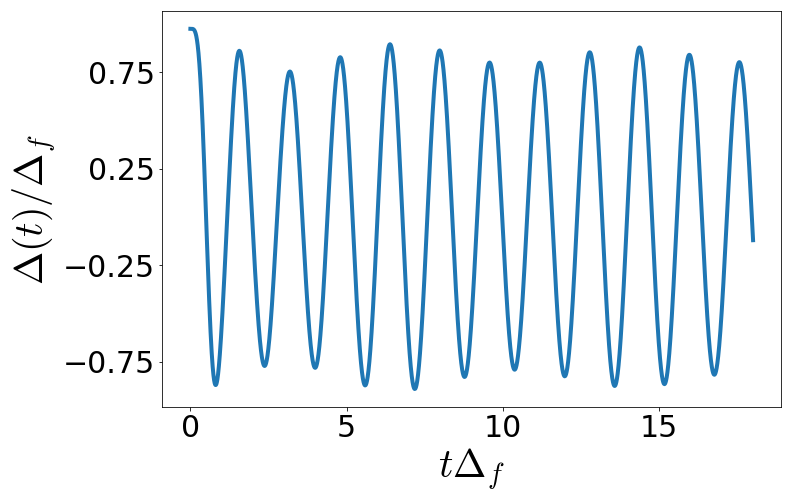}}
    \subfloat[ \label{fig:three_tanh_F}]{\includegraphics[width=.5\columnwidth]{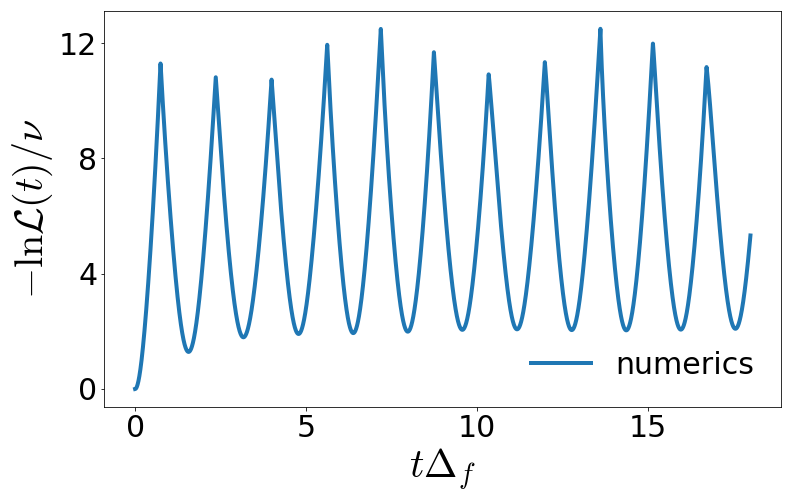}} \\
    \subfloat[ \label{fig:three_tanh_D_lt}]{\includegraphics[width=.5\columnwidth]{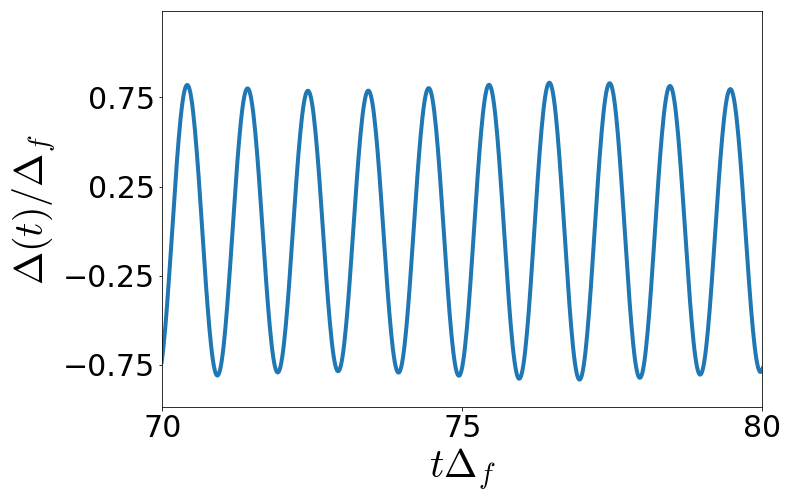}}
    \subfloat[ \label{fig:three_tanh_F_lt}]{\includegraphics[width=.5\columnwidth]{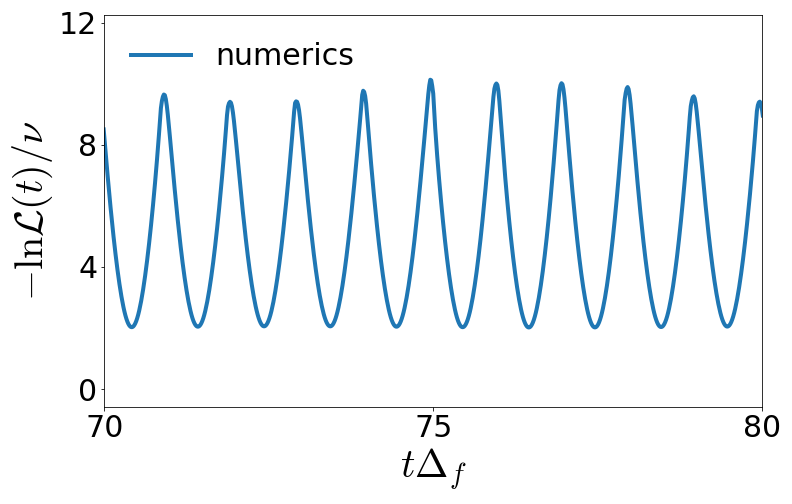}}
    \caption{Early [(a) and (b)] and late time [(c) and (d)] evolution of the order parameter $\Delta(t)$ and log of the Loschmidt echo
    with the $s$-wave BCS Hamiltonian  starting from the initial state~\re{eq:three_tanh} with $d=5$.  
    The initial spin distribution resembles those for unstable stationary states that produce the normal 2-soliton and single anomalous soliton,  except now it is a fully analytic function of $\xi_\p$. However, there are still DQPTs because  $s_\p^z(0)$   changes sign three times  creating zeros in the Cooper pair distribution shown in Fig.~\ref{fig:gammas}. Note also that the DQPTs are transient and disappear at late times.}
        \label{3tanhfig}
\end{figure}

\begin{figure}[hbt]
    \centering
    \subfloat[\label{fig:one_tanh_D}]{\includegraphics[width=.5\columnwidth]{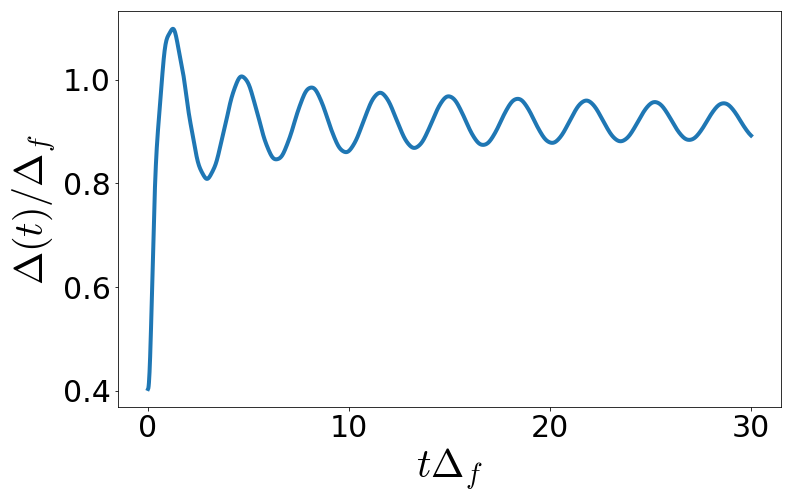}}
    \subfloat[\label{fig:one_tanh_F}]{\includegraphics[width=.5\columnwidth]{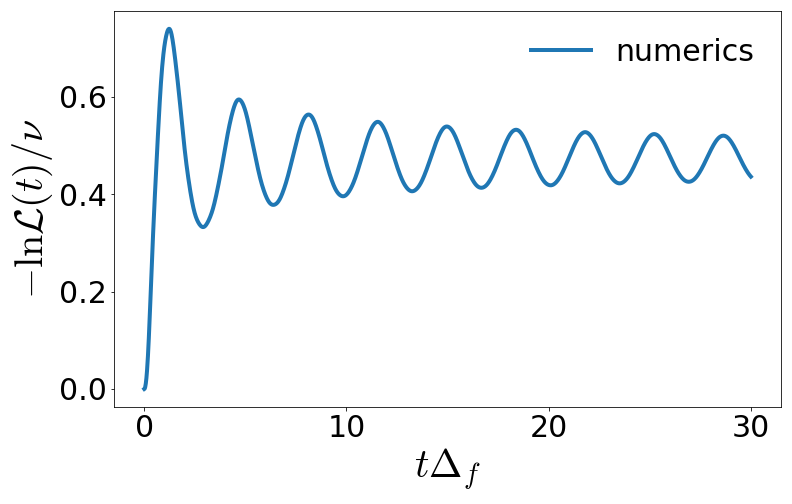}}
    \caption{ (a) Order parameter $\Delta(t)$ and (b) log of the Loschmidt echo for the time evolution with the $s$-wave BCS Hamiltonian   starting from the initial state~\re{eq:one_tanh}.   The pseudospin distribution  $\vec{s}_\p(t)$ resembles the  ground state one in that it is continuous and  $s_\p^z(0)$  changes sign only once. As a result there are no zeros in the Cooper pair distribution [see Fig.~\ref{fig:gammas}] and no    DQPTs. }
     \label{1tanhfig}
\end{figure}

\begin{figure}[hbt]
    \centering
    \includegraphics[width=.7\columnwidth]{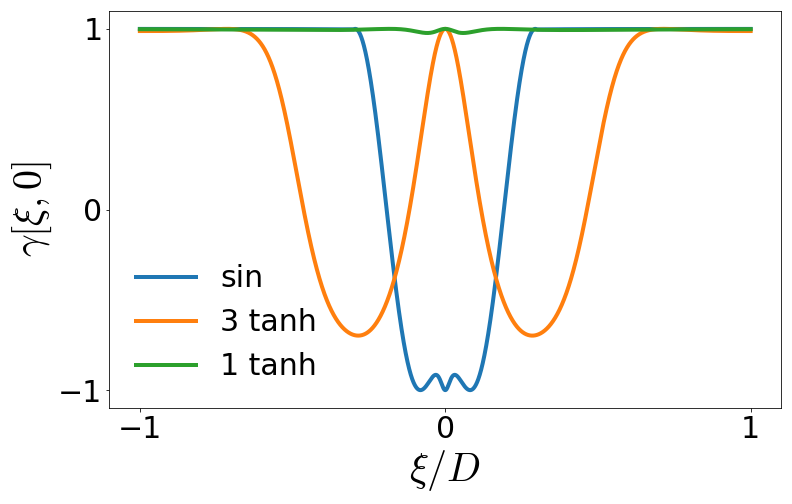}
    \caption{The Cooper pair distribution  $\gamma[\xi, t=0]$ for the three  initial conditions considered in this section. 
      By definition $\gamma[\xi, t]$ is the cosine of the angle between the spin $\vec{s}_\p$ and its effective field $\vec{B}_\p$. This
    plot further supports our conjecture that the existence of zeros in the Cooper pair distribution is a necessary condition for DQPTs.
     Specifically, we see that `sin' and `3 tanh' initial states for which $\gamma[\xi, 0]$ crosses through zero produce DQPTs, while
    the `1 tanh' initial state for which $\gamma[\xi, 0]$ has no zeros does not. The   order parameter $\Delta(t)$ and log of the Loschmidt echo for `sin', `3 tanh' and `1 tanh' initial conditions are shown in Figs.~\ref{sinfig}, \ref{3tanhfig} and \ref{1tanhfig}, respectively. 
   }
    \label{fig:gammas}
\end{figure}
 
 All our test initial states are particle-hole symmetric and have $s^z_{\vec{p}}\to\mp 1/2$ for $\xi_{\vec{p}}\to \pm\infty$  as well as $s^y_{\vec{p}}=0$ which are features of the states considered so far. The second property is a physical requirement that states far above [below] the Fermi energy be empty [occupied].
First we consider the following continuous spin distribution,
\beg
\begin{split}
2s^z_{\vec{p}}= \sin{\left[\frac{2\pi \xi_{\vec{p}}}{L}\right]}, \, 2s^x_{\vec{p}}= \cos{\left[\frac{2\pi \xi_{\vec{p}}}{L}\right]}, \quad |\xi_{\vec{p}}|<\frac{3L}{4},\\
s^x_{\vec{p}}=0,\, 2 s^z_{\vec{p}}=- \text{sgn}[\xi_{\vec{p}}],\quad|\xi_{\vec{p}}|>\frac{3L}{4},
\end{split}
\label{eq:sin_cos}
\en
where $L$ is a parameter such that $\delta\ll L<D$.   We numerically integrate the equations of motion \re{spinEOM} from this initial condition for $n=5\times 10^5$ spins, $L=4$, and coupling $g=g_f$ that corresponds to the ground state gap $\Delta_f=1.5$. From  Fig.~\ref{fig:sin_cos_F} we see that DQPTs are present for this distribution despite the lack of discontinuities in the initial state. Upon examining the Cooper pair distribution~\re{CPDF}, we find that it contains 4 zeros symmetric with respect to the Fermi energy, see  Fig.~\ref{fig:gammas}. 

The reason for the existence of the zeros is that $s_\p^z$ changes from $1/2$ to $-1/2$ and then to 0 as 
$\xi_\p$ goes from $-\infty$ to 0. As a result the angle the spin $\vec{s}_\p$ makes with the $z$-axis increases from 0 to $3\pi/2$, while the angle between $-\vec{B}_\p$ and the $z$-axis increases from 0 to only $\pi/2$. It is then inevitable that at some point $\vec{s}_\p$ must become perpendicular to $\vec{B}_\p$.  Generally, we can expect zeros in the Cooper pair distribution when $s_\p^z$ changes sign more than once reaching sufficiently negative values for $\xi_\p<0$  or sufficiently large positive values for $\xi_\p>0$.

The above state contains no discontinuities but is non analytic at $\xi=3L/4$ and so we consider now the following smooth distribution
\beg 
\begin{split}
s^z_{\vec{p}}&=\frac{\tanh{ \xi_{\vec{p}}}- \tanh{( \xi_{\vec{p}}-d)}-\tanh{( \xi_{\vec{p}}+d)}}{2},\\
s_{\vec{p}}^x&=\sqrt{\frac{1}{4}-(s_{\vec{p}}^z)^2},\quad s^y_{\vec{p}}=0,
\end{split}
\label{eq:three_tanh}
\en 
where  $d$ is an arbitrary constant. As always, $\xi_\p\in [-D, D]$ and we set the units so that $D=10$. The profile of $s_\p^z$ for this distribution resembles  that for both the normal 2-soliton and single anomalous soliton [see Sect.~\ref{solitonsec}], except it is a completely analytic function of $\xi_\p$.  As shown in  Fig.~\ref{fig:three_tanh_F} DQPTs do occur but they are transient, disappearing at late times as the cusps are smoothed out. We note also that the disappearance of the DQPTs does not coincide with any noticeable change in the behavior of the order parameter. As shown in  Fig.~\ref{fig:gammas} these DQPTs can be understood by the presence of zeros in the initial Cooper pair distribution function. Similar to the previous example, the zeros in turn are a consequence 
of strong oscillations around 0 in $s_\p^z$ as a function of $\xi_\p$ in the initial state.

Lastly we study the dynamics emerging from the state
\begin{equation}
s^z_{\vec{p}}=\frac{ \tanh \xi_\p}{2},\quad s^x_{\vec{p}}=\frac{1}{2\cosh\xi_\p},\quad s^y_{\vec{p}}=0.
\label{eq:one_tanh}
\end{equation}
Again this is a smooth distribution, this time approximating the superconducting ground state. We run this initial condition as well as \eqref{eq:three_tanh} with $n=5\times 10^5$ spins and $\Delta_f=1.8$. In this case, see  Figs.~\ref{fig:one_tanh_F} and ~\ref{fig:gammas}, the are no DQPTs which occur and $\gamma[\xi_{\vec{p}}, t=0]$ has no zeros.  Note also
that~\eqref{eq:one_tanh} is the $d\to0$ limit of~\eqref{eq:three_tanh}. A numerical check shows that the zeros and DQPTs disappear
already  at a finite value of $d$ while $s_\p^z$ still changes sign 3 times. Therefore, it is not sufficient that $s_\p^z$  change sign more than once, but also  deviations from zero following the sign changes must be   sufficiently  large. 

From these three cases we see again that the existence of DQPTs is linked to the appearance of zeros in the Cooper pair distribution  but which cannot necessarily be linked to any long time behavior of the order parameter.  We also conclude that discontinuities 
or any other non-analyticities in the spin distribution are not necessary for having DQPTs. Moreover, DQPTs occur for 
initial states where the fermion occupancy $n_\p=2s^z_\p+1$  oscillates strongly around $n_\p=1$ and which are otherwise arbitrary.

\section{Quenches in a P+iP superfluid }
\label{pip}

We have shown that the presence or absence of DQPTs in the quench dynamics of the  BCS superconductor cannot be used to infer any behavior of the order parameter at long time. We can then ask if there are  any other properties which it can provide information for. To answer this we briefly look at a different model, the 2D topological $p$-wave  superfluid~\cite{GURARIE20072}
\beg \label{fullpwave} 
 \hat H = \sum_{\p} \frac{p^2}{2m} \ad_\p \a_\p -  2G{\sum_{\q, \k}}^\prime \q\cdot \k \, \ad_{ \q} \ad_{ - \q} \a_{ - \k} \a_{\k},
\en
where $\ad_\p$ and $\a_\p$ are creation and annihilation operators  of spinless fermions of mass $m$ with 2D momentum $\p$
and $G>0$ is their interaction strength. The prime over the second summation indicates that it is over only those $\q$ and $\k$ that satisfy $q_x>0$ and $k_x>0$ and $p$ is the magnitude of vector $\p$.

An important difference between this and the $s$-wave BCS models is that here there is a quantum phase transition   between topologically non-trivial   BCS phase and topologically trivial 
  BEC phases. This transition occurs at finite coupling $G$ at the point where the chemical potential $\mu_i$ vanishes and
excitations become gapless. Therefore, in this model we will be able to study superfluid $\to$ superfluid   quenches across the quantum critical point.

The ground state of this Hamiltonian is a $p_x+ip_y$ superfluid~~\cite{GURARIE20072}. One can show that the  $p_x+ip_y$  symmetry is preserved by the dynamics and as a result the time evolution with the Hamiltonian~\re{fullpwave} starting from a $p_x+ip_y$ state is identical to that with a   Hamiltonian~\cite{FDGY} 
\beg \label{chiralpwave} 
\hat  H = \sum_{\p} \frac{p^2}{2m} \ad_\p \a_\p -  G{\sum_{\p, \k}}^\prime p k \, \ad_{ \p} \ad_{ - \p} \a_{ - \k} \a_{\k},
\en
which is quantum integrable~\cite{Dunning_2010}.
 For the same reasons  as for the $s$-wave BCS model, we expect the mean field description to become exact in the thermodynamic limit. The system is then in a product state at all times,
 \begin{eqnarray}\label{psitpwave}
\ket{\Psi_\mathrm{BCS}(t)}=\prod_{\p}\left[u^*_{\p}(t)+v^*_{\p}(t)c^\dag_{\p }c^\dag_{-\p}\right]\ket{0},
\end{eqnarray}
which time evolves with the mean field Hamiltonian
\beg
\begin{split}
\hat H(t)=&\sum_{\p}\ \frac{p^2}{2m} c^\dag_{\p}c_{\p}-\Biggl(\Delta(t)\sum_{\p}p c^\dag_{\p}c^\dag_{-\p} +\mathrm{h.c.}\Biggr),\\
&\Delta(t)=G \sum_{\p} p \langle  c_{-\p } c_{\p }\rangle,
\end{split}
\label{pmf}
\en
where $\Delta(t)$ is the time dependent $p$-wave order parameter. In the ground state $\Delta(t)=\Delta_0 e^{-2i\mu t}$, where $\Delta_0$ is a constant and  $\mu$ is the chemical potential.

 Classical pseudospins are defined as
\begin{align}\label{Pseudospins}
\begin{aligned}
	&2s_{\vex{k}}^z 
	=\langle c^\dagger_{\vex{k}}c_{\vex{k}}+c^\dagger_{-\vex{k}}c_{-\vex{k}}\rangle-1=|v_\p|^2-|u_\p|^2,\\
	&  s_{\vex{k}}^-=\langle c_{-\vex{k}} c_{\vex{k}}\rangle=u_\p v_\p^*,\quad s_{\vex{k}}^+=\left(s_{\vex{k}}^-\right)^*,
\end{aligned}
\end{align}
where as before the operators are in the Schr\"odinger picture, while quantum expectation values are with respect to the time-dependent state of the system~\re{psitpwave}. They evolve with the classical Hamiltonian~\cite{FDGY} 
\begin{eqnarray}\label{pwaveH}
H_c=\sum_{\vec{p}}2\eps_{\vec{p}}s^z_{\vec{p}}-G\sum_{\vec{k}, \vec{p}}\sqrt{\eps_{\vec{k}}\eps_{\vec{p}}}s^+_{\vec{k}}s^-_{\vec{p}}
\end{eqnarray}
where $\eps_{\vec{q}}\equiv q^2$ and we set the units of mass so that $2m=1$.  More specifically,  the spin equations of motion are
\beg
\begin{split}
&\dot{\vec{s}}_{\vec{p}}=\vec{B}_{\vec{p}}\times \vec{s}_{\vec{p}},\\
&\vec{B}_{\vec{p}}=-2\sqrt{\eps_\p}(\Delta_x \hat x+\Delta_y\hat y)+2\eps_\p\hat z,
\end{split}
\en
where
\begin{eqnarray}
\Delta\equiv\Delta(t)\equiv\Delta_x-i\Delta_y=G\sum_{\vec{q}}\sqrt{\eps_{\vec{q}}}s^-_{\vec{q}},
\label{Deltaspinsp}
\end{eqnarray}
is the   $p$-wave order parameter now written in terms of classical spins.  As before the Cooper pair distribution is defined as
\begin{equation}
\label{CPDFp}
\gamma[\eps_{\vec{p}}, t] =\frac{2{\vec{s}_{\vec{p}}} \cdot\vec{B}_{\vec{p}}}{|\vec{B}_{\vec{p}}|},
\end{equation}
and  the Loschmidt echo in terms of pseudospins reads
\begin{eqnarray}\label{SpinEchop}
\mathcal{L}(t)=\prod_{\vec{p}}\left[\frac{1}{2}+2\vec{s}_{\vec{p}}(0)\cdot\vec{s}_{\vec{p}}(t)\right].
\end{eqnarray}
The ground state of the 2D $p$-wave superfluid  is either in a topologically non-trivial weak-pairing BCS phase or topologically trivial 
strong-pairing BEC phase. The topological quantum phase transition between the two  occurs as a function of the coupling
strength $G$ at the point where the  chemical potential vanishes, $\mu_i=0$. At this point the bulk quasiparticle spectrum develops a massless Dirac node at $p=0$. These two phases can be distinguished by a bulk topological invariant. One formulation defines this invariant as the winding number $Q$ of the pseudospin texture $\vec{s}_{\vec{p}}$. 

Another approach is to define a topological invariant $W$ in terms of the retarded
single-particle Green's functions, which is equivalent to the winding of the effective magnetic field 
\beg
\vec{B}^{\mathrm{eff}}_{\vec{p}}=
\vec{B}_{\vec{p}}-2\mu_i\hat z.
\label{beff}
\en
 In equilibrium $Q=W,$
because $\vec{s}_{\vec{p}}$ is parallel to $\vec{B}^{\mathrm{eff}}_{\vec{p}}$. In particular, in the non-trivial BCS phase $W=Q=1$, while $W=Q=0$ in the trivial BEC phase. Out of equilibrium $\vec{s}_{\vec{p}}$ is no longer aligned with $\vec{B}^{\mathrm{eff}}_{\vec{p}}$ and the two winding numbers do not have to coincide. Moreover, it is $W$ and not $Q$ which indicates the presence of zero energy Majorana edge modes in a finite sample.

 The solution for the dynamics of the $p_x+ip_y$ superfluid in the thermodynamic limit can be determined with same methods as for the $s$-wave model. The resulting phase diagram~\cite{FDGY} displays the same three main nonequilibrium  phases    where the amplitude of the order parameter either vanishes [Phase I], asymptotes to a constant [Phase II] or persistently oscillates [Phase III]. There are however some new features related to nonequilibrium topology. As mentioned above, the two topological indices are not equivalent out of equilibrium. The winding of the pseudospin texture $Q$ turns out to be  a constant of motion because the spin
 distribution is pinned at $p=0$ and $p=\infty$.
 The winding of the effective field $W$ however, while still quantized at late times, is not conserved and in the long time limit we may have that $Q\neq W$ if the system is quenched across a quantum critical point. This subdivides the phase diagram into different regions depending upon the values of $Q,W$ in the asymptotic state.  Moreover, it is $W$ and not $Q$ which indicates the presence of zero energy Majorana edge modes in a finite sample thus allowing for these states to appear even when quenched form a trivial state. 
 
 \begin{figure}
    \centering
    \subfloat[\label{fig:chiral_gamma_dqpt}]{\includegraphics[width=.5\columnwidth]{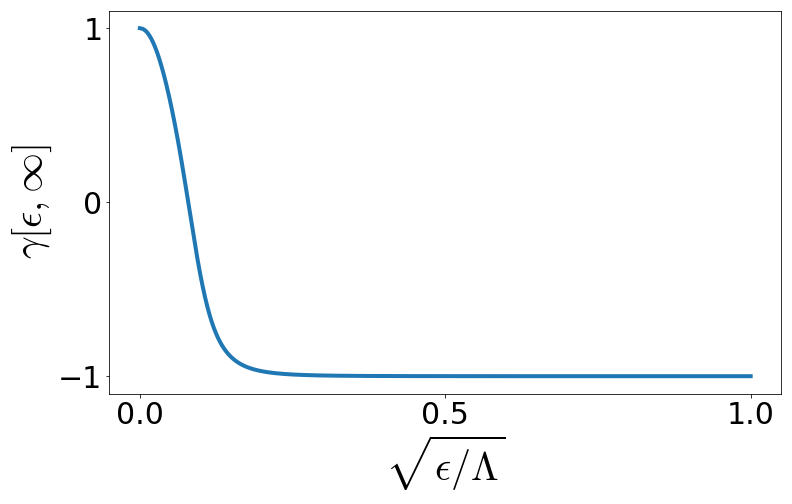}}
    \subfloat[\label{fig:chiral_D_dqpt}]{\includegraphics[width=.5\columnwidth]{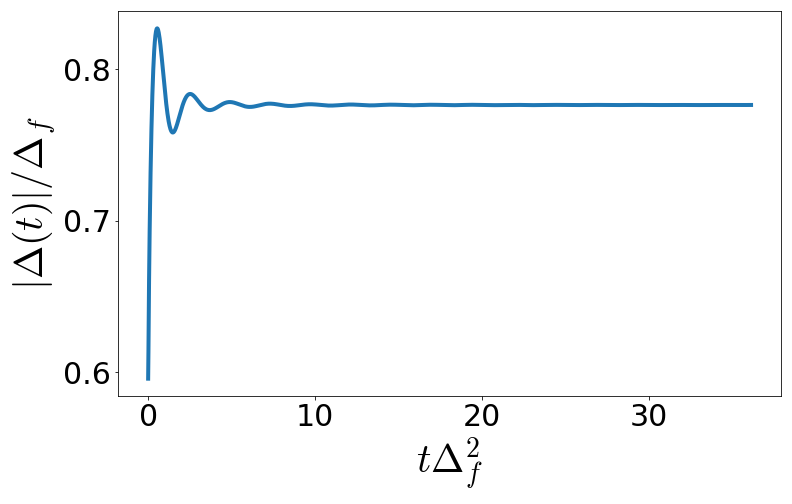}} \\
    \subfloat[\label{fig:chiral_F_dqpt_norotate}]{\includegraphics[width=.5\columnwidth]{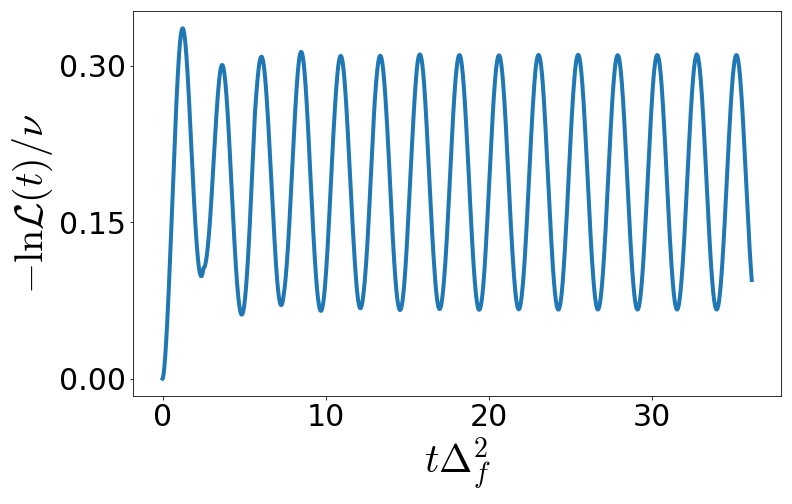}}
    \subfloat[\label{fig:chiral_F_dqpt_rotate}]{\includegraphics[width=.5\columnwidth]{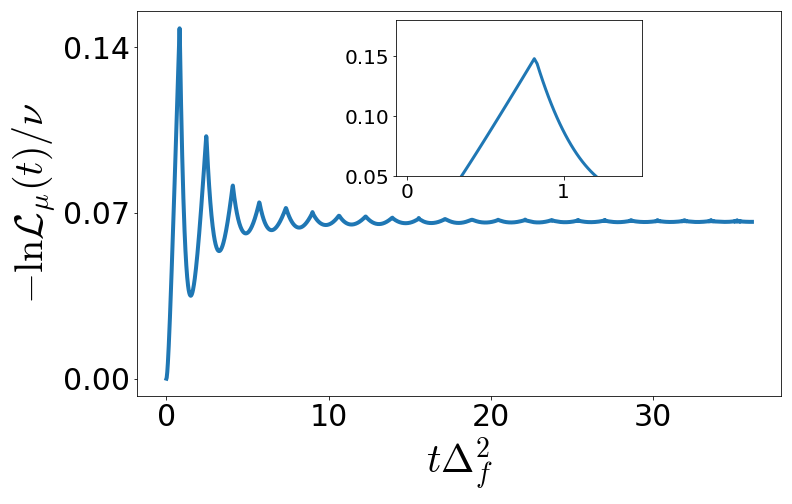}}
    \caption{Loschmidt echo detects the disappearance of Majorana modes after a quench, but  only after a particular  global $U(1)$  rotation of the superfluid phase. This shows that, unlike physical properties of an isolated superfluid, the existence of DQPTs is not invariant under such rotations. Here we show an interaction quench of a 2D $p_x+ip_y$ superfluid, such that (b) the magnitude of the order parameter $|\Delta(t)|$ asymptotes to a nonzero constant and the winding number $W$ equals 1 before and 0  
    long after the quench.   Even though
    (a) the Cooper pair distribution has   a zero in the steady state,   (c)  there are no DQPTs in the Loschmidt echo.  Killing the phase of $\Delta(t)$  with an appropriate $U(1)$ rotation,  (d)  brings about DQPTs whose amplitudes decay, see inset for a close up of the first DQPT.  }
    \label{fig:chiral_dqpt}
\end{figure}

\begin{figure}
    \centering
    \subfloat[\label{fig:chiral_gamma_nodqpt}]{\includegraphics[width=.5\columnwidth]{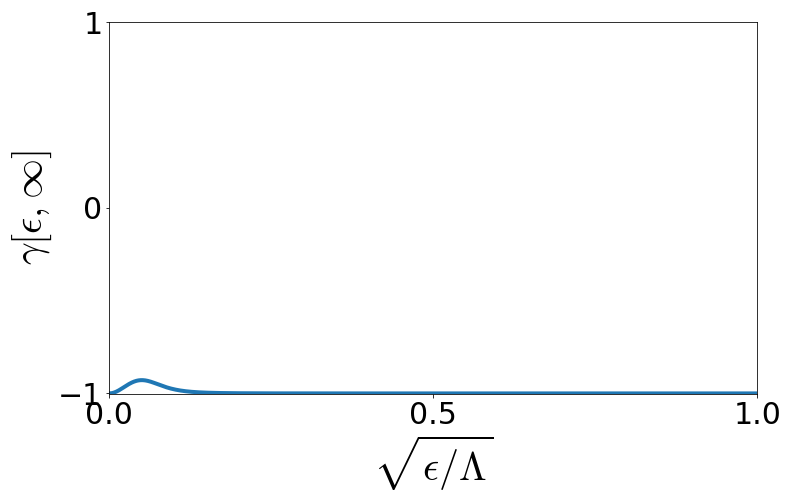}}
    \subfloat[\label{fig:chiral_D_nodqpt}]{\includegraphics[width=.5\columnwidth]{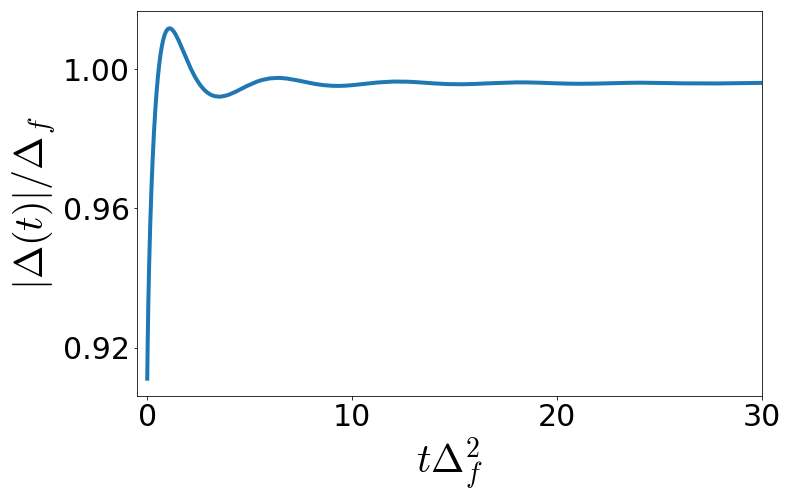}} \\
    \subfloat[\label{fig:chiral_F_nodqpt_norotate}]{\includegraphics[width=.5\columnwidth]{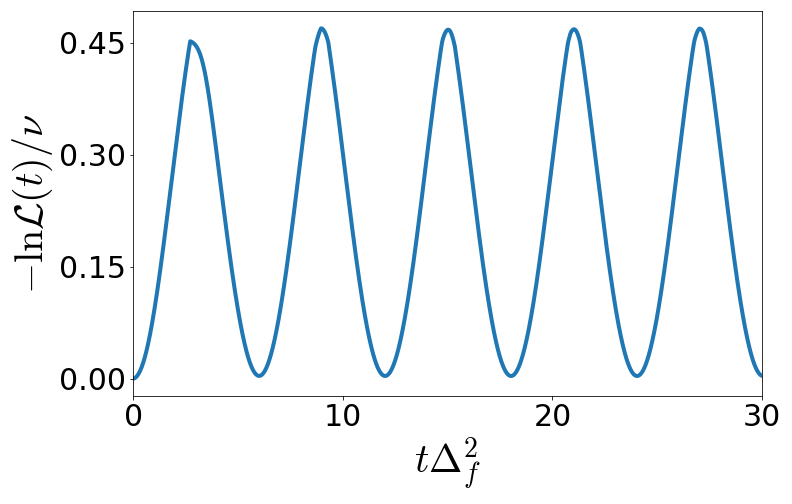}}
    \subfloat[\label{fig:chiral_F_nodqpt_rotate}]{\includegraphics[width=.5\columnwidth]{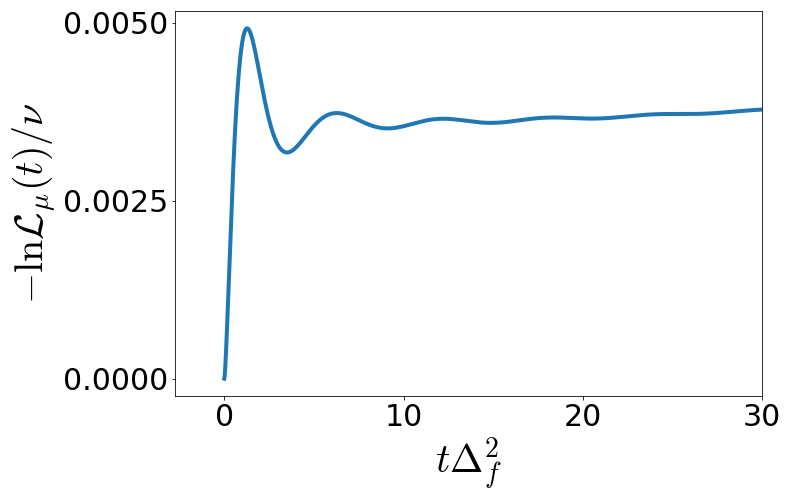}}
    \caption{As in Fig.~\ref{fig:chiral_dqpt}, (b) the magnitude of the order parameter $|\Delta(t)|$ asymptotes to a nonzero constant after an interaction quench of a 2D $p_x+ip_y$ superfluid, but here $W=1$ and  Majorana edge modes are present both before and after the quench. Now
    (a) the Cooper pair distribution function has no zeros  and the Loschmidt echo shows no DQPTs both (c) in the original reference frame and (d) after a time-dependent $U(1)$ phase rotation that kills the phase of $\Delta(t)$.}
    \label{fig:chiral_nodqpt}
\end{figure}

 This dramatic difference between equilibrium and nonequilibrium topologies   arises from the time dependence of the phase of $\Delta(t)$, which is now a dynamic quantity due to the  absence of the particle-hole symmetry in the $p$-wave Hamiltonian [see the
 discussion around \eqref{phsym}]. In particular, in Phase II we have 
 $\Delta(t)\to \Delta_\infty e^{-2i\mu_\infty t}$. The effective chemical potential $\mu_\infty$ can differ in sign from the chemical potential $\mu_i$ of the pre-quench state. This allows the winding $W$ of the effective field~\re{beff}, where
 $\mu_i$ is now replaced with $\mu_\infty$, to change leading to $W\ne Q$. In contrast to the $s$-wave model considered above, the
 pair distribution   $\gamma[\eps_{\vec{p}},t\to\infty]$ may now exhibit a number of permanent zeros in Phase II. Interestingly, it was realized~\cite{FDGY} that the parity of these is linked to the change in $W.$ For an even number,   $W=Q$ in the asymptotic state and thus the initial and steady states have the same topological properties. On the other hand, when $\gamma[\eps_{\vec{p}},t\to\infty]$ has an odd number of zeros, $W\neq Q$ and the topology has changed as a result of the quench.
 
 In our study  of the $s$-wave BCS superfluid we saw that DQPTs were linked to the zeros of the Cooper pair distribution. Therefore, it makes sense to ask whether DQPTs are present in Phase II of nonequilibrium $p_x+ip_y$ superfluid and whether they carry any information about the topology of this steady  state of the system. These questions can in principle be answered analytically as the steady state wavefunction is known exactly. However, for our purposes it is simpler to address them numerically. 
 To this end, we prepare
$n = 5\times 10^3$  spins in the ground state for a certain initial coupling $G_i$,
 \beg
 s_\p^-=\frac{\sqrt{\eps_\p} \Delta_i}{2E_\p},\quad s_\p^z=-\frac{ \eps_\p- \mu_i}{2E_\p},
 \en
 where $E_\p=\sqrt{(\eps_\p -\mu_i)^2+\eps_\p\Delta_i^2}$ is the quasiparticle energy. The ground state gap $\Delta_i$ and
 chemical potential $\mu_i$  are related to the initial coupling $G_i$ and the average fermion number $N$ through the self-consistency 
  and chemical potential  equations,
 \beg
 \Delta_i=G_i\sum_\p s_\p^-,\quad N=\sum_\p(2s_\p^z+1).
 \label{pcp}
 \en
  We then quench the coupling $G_i\to G_f$ and as before characterize each such quench by $\Delta_i$ and the value of the gap $\Delta_f$ in the ground state with coupling $G_f$.
 In numerical simulations, we choose the 2D fermion number density   $n_\mathrm{f}=0.825$ and the ultraviolet cutoff 
 $\Lambda= 100\pi n_\mathrm{f}$. See~\cite{FDGY}  for the details on how to determine $W$ and $Q$  and relate
 $n_\mathrm{f}$ and $\Delta_{i, f}$ to $G_{i,f}$. With these parameters the quantum critical point is at $\Delta_\mathrm{QCP}=1.53$
 
 Let us analyze two interaction quenches. One is across the quantum critical point, $\Delta_i=1.0<\Delta_\mathrm{QCP}$ and $\Delta_f=1.9>\Delta_\mathrm{QCP}$. We show the results for this quench in Fig.~\ref{fig:chiral_dqpt}. 
  The winding numbers in this case are $Q=W=1$ before and $Q=1, W=0$ after the quench, respectively. 
  The second quench shown in Fig.~ \ref{fig:chiral_nodqpt} is within the BCS phase. Here both $\Delta_i=1.21$ and $\Delta_f=1.35$
  are smaller than $\Delta_\mathrm{QCP}$. This quench is similar to the quenches of the $s$-wave BCS superconductor we considered before. Now $Q=W=1$ before and after the quench. 
 
 We see immediately  from Figs.~\ref{fig:chiral_gamma_dqpt} and \ref{fig:chiral_F_dqpt_norotate} and Figs.~\ref{fig:chiral_gamma_nodqpt} and \ref{fig:chiral_F_nodqpt_norotate} that  whether  $\gamma[\eps_\p,\infty]$ has zeros or not,
 there are no DQPTs in a seeming contradiction to our conclusions for the $s$-wave BCS model. The reason is that now the order parameter, $\Delta(t)\equiv \Delta_x+i\Delta_y= \Delta_\infty e^{-2i\mu_\infty t}$, winds around the origin in the $xy$-plane with frequency $2\mu_\infty$ at large times.
 This did not happen in the $s$-wave case, because we imposed particle-hole symmetry, which ensured $\mu_\infty=0$.  We can eliminate this overall rotation around the $z$-axis   by moving to the rotating frame, 
 \beg
 s^-_\p= \tilde s^-_\p e^{-2i\mu_\infty t},\quad s_\p^z=\tilde s_\p^z.
 \label{srot}
 \en 
  In the new reference frame the order parameter is constant in Phase II steady state,  $\tilde\Delta(t)=\Delta_\infty$.     Spins  now presses around static magnetic fields as in the noninteracting model we considered  in Sect.~\ref{nointsect}. Every zero in the Cooper distribution will therefore generate a  periodic sequence of DQPTs provided we evaluate
 the Loschmidt echo in the rotating frame.

 The Cooper pair distribution~\re{CPDFp} is   invariant with respect to  any such time-dependent U(1) phase rotation.  In contrast, the  Loschmidt echo  is \textit{not invariant}, because it involves spin-spin correlator at unequal  times. 
 Evaluated in the rotating frame, the echo~\re{SpinEchop} becomes
\begin{equation}
\begin{array}{ll}
 \dis \mathcal{ L}_{\mu}(t)= \prod_{\vec{p}}\left[\frac{1}{2}+2 \vec{\tilde s}_{\vec{p}}(0)\cdot\vec{\tilde s}_{\vec{p}}(t)\right]=\prod_{\vec{p}}\biggl[\frac{1}{2}+2s^z_{\vec{p}}(0) s^z_{\vec{p}}(t) \biggr.\\
\dis \quad \biggl. +
s^+_{\vec{p}}(0) s^-_{\vec{p}}(t) e^{2i\mu_\infty t} +s^-_{\vec{p}}(0) s^+_{\vec{p}}(t)  e^{-2i\mu_\infty t} \biggr]\ne \mathcal{L}(t). 
\end{array}
\label{rotecho}
\end{equation}
 As anticipated, the connection between quenches across the quantum critical point, zeros of the Cooper pair distribution and DQPTs is restored if we use $\mathcal{L}_\mu(t)$ instead of $\mathcal{L}(t)$.  Indeed, note a  periodically repeating DQPT in Fig.~\ref{fig:chiral_F_dqpt_rotate} when $\gamma[\eps_{\vec{p}}, \infty]$ has a zero and no DQPTs in Fig.~\ref{fig:chiral_F_nodqpt_rotate} when it does not. The absence of DQPTs in the latter case immediately tells us that
$W=Q$ in the steady state and that the initial state and the ground state of the final Hamiltonian belong to the same equilibrium phase, BCS
or BEC. A periodic sequence of DQPTs in Fig.~\ref{fig:chiral_F_dqpt_rotate} is  consistent with one zero implying that in the steady state the value of $W$  is different from that in the initial state and $W\ne Q$. This is therefore a quench  across the quantum critical point. 

It should be mentioned that the Cooper pair distribution can have more than one zero both for quenches across the quantum critical point and for quenches within the same equilibrium phase~\cite{FDGY} . Therefore, it is possible to have DQPTs  in both cases. It is only the parity of the number of zeros that depends on the type of the quench.   Each zero generates its own periodic sequence of  DQPT, which we count as one `independent' DQPT. The number
of independent DQPTs is odd for for quenches across the quantum critical point and even otherwise.
 Odd number of DQPTs also indicates that the winding number $W$ in the steady state is different from that in the initial state, i.e., Majorana modes have either emerged or disappeared as a result of the quench. As with the $s$-wave case however DQPTs cannot tell us about the behaviour of the order parameter or in which of the three nonequilibrium phases we are.
 
 \section{Grand canonical Loschmidt echo}
 \label{grand}
 
 We saw that we had to redefine the Loschmidt echo $\mathcal{ L}(t)\to\mathcal{ L}_\mu(t)$ to preserve the link between DQPTs and quenches across the quantum critical point in the $p_x+ip_y$ superfluid. Let us formulate the new definition   more generally.  
   Consider the following transformation of the Bogolioubov amplitudes, 
 \beg
 u_\p(t)=  \tilde u_\p(t),\quad  v_\p(t)= \tilde v_\p(t)e^{2i\mu t},
 \en
For $\mu=\mu_\infty$ this is equivalent to the transformation to the rotating frame in \eqref{srot}. We had to replace the BCS wavefunction~\re{psitpwave} with
 \beg
 | \widetilde{\Psi }(t)\rangle=\prod_{\p}\left[\tilde u^*_{\p}(t)+\tilde v^*_{\p}(t)c^\dag_{\p }c^\dag_{-\p}\right]\ket{0}.
 \en
This is a unitary operation because it preserves the norm.  Alternatively, we could rotate the phase of the fermion creation and annihilation operators, $c_{\vec p} = \tilde c_{\vec p}e^{-i\mu_\infty t}$ and $c^\dagger_{\vec p} = \tilde c^\dag_{\vec p} e^{i\mu_\infty t}$. 
Physical properties of an isolated nonequilibrium  superfluid, such as the condensate fraction, radio-frequency (RF) absorption spectrum of  a paired unltracold gas or the optical conductivity  of a metallic superconductor~\cite{DzeroyuzbashAltColeman,PhysRevA.92.053620,PhysRevB.95.104507}   are  invariant under such global U(1)
phase rotations. The Loschmidt echo as defined in \eqref{echo} is not invariant as we saw in the previous section.

The operator canonically conjugated to the phase is half the
  total fermion number operator $\hat N/2$. It follows that to translate the phase by $-2\mu t$ we need to apply $\exp[i\mu \hat N t]$ to the wavefunction, i.e., 
  \beg
  | \widetilde{\Psi }(t)\rangle  = e^{i\mu \hat N t}\ket{  \Psi (t)}=e^{-i(\hat H -\mu \hat N) t}\ket{  \Psi_i},
  \en
We therefore need to redefine the Loschmidt amplitude as $\langle \Psi_i |\Psi(t)\rangle \to \langle \Psi_i |\widetilde{\Psi}(t)\rangle = \matrixel{\Psi_i}{e^{-i(\hat H -\mu \hat N) t}}{\Psi_i}$. For the pedagogical purposes we assumed in this paragraph only that the Hamiltonian 
$\hat H$ is time-independent and commutes with $\hat N$. We are now ready to lift this assumption.

    We define  the \textit{grand canonical Loschmidt amplitude} for an arbitrary 
time-dependent Hamiltonian without assuming particle number conservation as
\beg
\begin{split}
\mathcal{G}_\mu(t)&=\matrixel{\Psi_i}{  U_\mu (t) }{\Psi_i}=\langle \Psi_i | \Psi_\mu(t)\rangle,\\
 U_\mu (t)&=\mathsf{T}\exp\left\{ -i \int_{t_i}^t  dt\left[\hat H(t)-\mu(t) \hat N\right]\right\},
 \end{split}
 \label{newecho}
\en
where $\mathsf{T}$ is the time ordering operator. The \textit{grand canonical  Loschmidt echo}  is $\mathcal{L}_\mu(t)=|\mathcal{G}_\mu(t)|^2$. The old, canonical  Loschmidt amplitude given by~\eqref{echo}  is a particular $\mu(t)\equiv0$  case of the grand canonical one.
  It  can also be seen that the two definitions are equivalent if the Hamiltonian commutes with $\hat N$ and the initial state is a particle number eigenstate. The mean-field Hamiltonians~\re{mfH}
and \re{pmf} we employed to study the $s$ and $p$-wave superfluid dynamics do not commute with $\hat N$ and the BCS wavefunction does not have a definite particle number. For them the difference between canonical and grand canonical Loschmidt echos is essential unless $\mu(t)\equiv 0$ by symmetry. In the $p_x+ip_y$ superfluid $\mu(t)=\mu_\infty$ and in the particle-hole symmetric  $s$-wave BCS superconductor   $\mu(t)=0$.   

It is important to distinguish  two  Loschmidt echos in this discussion. Both are given by~\eqref{echo}. One, $\mathcal{L}_\mathrm{ex}(t)$, is the exact  echo evaluated for the original particle number conserving Hamiltonian~\re{H} or~\re{fullpwave}  with $|\Psi_i\rangle$ being a particle number eigenstate. The other, $\mathcal{L}_\mathrm{mf}(t)$,
is evaluated with the corresponding mean-field Hamiltonian and a BCS-like $|\Psi_i\rangle$. In this notation, our claim is that in the thermodynamic limit $\mathcal{L}_\mu(t)=\mathcal{L}_\mathrm{ex}(t)\ne\mathcal{L}_\mathrm{mf}(t)$.

Let us investigate the properties of $\mathcal{L}_\mu(t)$ in the context of superfluid dynamics.  The operator $U_\mu(t)$ in \eqref{newecho} describes the evolution with an effective Hamiltonian
$\hat H_\mu(t)=\hat H(t)-\mu(t) \hat N$. The addition of $-\mu(t) \hat N$  to the  $s$ and $p$-wave Hamiltonians given by Eqs.~\re{mfH} and \re{pmf} shifts the single particle energies by $-\mu(t)$.  This changes   the magnetic field, $\vec{B}_\p\to\vec{B}_\p-2\mu(t)\hat z$, in both cases.
We can undo the effect of this additional term if we move to the rotating frame as in~\eqref{srot}
 \beg
 s^-_{\p}=   s^-_{\p\mu} e^{-2i\chi(t)},\quad s_{\p}^z=  s_{\p\mu}^z,\quad \dot\chi(t)=\mu(t).
 \label{smu}
 \en 
 Here $\vec{s}_\p$ are the usual pseudospins which evolve with $\hat H(t)$ and determine the state of the system $|\Psi_\mathrm{BCS}(t)\rangle$. 
 Auxiliary spins  $\vec{s}_{\p\mu}$ evolve with $\hat H_\mu(t)$ and determine the grand canonical Loschmidt echo through the first equation in~\re{rotecho}, where now 
 $\vec{\tilde s}_{\p}(t)=\vec{s}_{\p\mu}(t)$. 
 
 The auxiliary collective field $\Delta_\mu(t)$ is defined in terms $s^-_{\p\mu}$ in the same way as $\Delta(t)$ is defined in terms of $s^-_{\p}$ in Eqs.~\re{Deltaspins} and \re{Deltaspinsp}. Therefore,
\beg
\Delta_\mu(t)=\Delta(t)  e^{2i\chi(t)}.
\en
 Our choices $\mu(t)=\mu_\infty$ for $p$-wave and $\mu(t)=0$ for $s$-wave make the phase of $\Delta_\mu(t)$ time independent, at least at large times. The field $\Delta_\mu(t)$ is therefore real up to a constant phase factor that can be absorbed into fermion creation and annihilation operators with no harm. Then, the mean-field interaction  in $\hat H_\mu(t)$ [the bracketed part of Eqs.~\re{mfH}
and \re{pmf} with $\Delta(t)\to\Delta_\mu(t)$] is particle-hole symmetric.  It is presently unclear if this property qualifies as a general criterion for determining $\mu(t)$, but  at least it  works in the examples we considered above.  Note in this regard that that the interaction term was particle-hole symmetric in the original Hamiltonians~\re{H} and~\re{fullpwave} before the mean-field decoupling.

By analogy with the grand canonical ensemble, it is tempting to attempt to fix
$\mu(t)$ with the help of the equation $N= \langle \Psi_\mu(t) |\hat N|\Psi_\mu(t)\rangle$, where $|\Psi_\mu(t)\rangle=U_\mu(t)|\Psi_i\rangle$ and $N$ is the average particle number in the initial state. This approach does not work, because $\langle \Psi_\mu(t) |\hat N|\Psi_\mu(t)\rangle$ is independent of $\mu(t)$ as evident from the second equations in~\re{smu} and~\re{pcp}, so any $\mu(t)$ satisfies this equation. Finally, note that the grand canonical Loschmidt echo is invariant under arbitrary time-dependent U(1) rotations of the
superfluid phase. Indeed, $e^{i\kappa \hat N} |\Psi_\mu(t)\rangle=|\Psi_{\mu'}(t)\rangle$ with $\mu' =\mu+\dot \kappa$, which is just a renaming of a variable $\mu\to \mu'$.

\section{Conclusions}

In this paper we have examined the relevance of non-analytic points of the  Loschmidt echo, DQPTs, for determining the long time dynamical phases of  fermionic  superfluids. Our work represents the most complete study of DQPTs undertaken in an interacting model, that we are aware of. We have examined the entire quantum quench phase diagram of the BCS model as well as    the dynamics emerging from unstable stationary states and more arbitrary initial conditions. We have also investigated   far from equilibrium topological 2D $p$-wave superfluid, where interaction quenches can change the winding number.  We have done this using explicit calculation supported by numerical simulation. The results  highlight several features of DQPTs which manifest in the interacting model and cast doubt on their usefulness as a predictor of dynamical phases. 

In general, we find no correlation between DQPTs and the long time dynamics of the interacting system. In particular, quenched BCS superconductors end up in one of the three distinct  steady states depending on the strength of the quench. These are true dynamical phases separated by second order phase transition lines. Yet there is not a single DQPT in the dynamics leading to any of these steady states.

We showed that the only necessary condition for having DQPTs in both  interacting and non-interacting cases is the existence of zeros in the Cooper pair distribution function. In the latter case, these zeros are constants of motion and DQPTs occur periodically. 
In an interacting system, zeros are time-dependent, DQPTs do not occur periodically and, moreover, are often transient. As a result they can be removed or induced by time translation of the initial state, which shows that the same steady state may be reached with or without DQPTs.

  We have seen that largely arbitrary initial states can produce DQPTs.  One scenario is to prepare a  state where the fermion occupation number $n(\xi)$  makes several large amplitude oscillations around its median value  as a function of the single-particle energy $\xi$. This creates zeros in the Cooper pair distribution, which in turn generate DQPTs.
In this case, $n(\xi)$ can be a continuous function. On the other hand, particle-hole symmetric initial states with maximal discontinuity
in $n(\xi)$ at the Fermi level also produce DQPTs. Here there is a permanent zero in the Cooper pair distribution at the Fermi level and as a consequence DQPTs can persist indefinitely. In particular, the ground state of the free Fermi gas falls into this class of initial states. 
The only overall physical principle that unites  initial states that lead to DQPTs is that they are distinct from the BCS ground state, e.g.,
show oscillations or discontinuities in $n(\xi)$. This principle is however not exclusive as there are equally many states similarly distinct from the ground state that do not produce any DQPTs, such as, for example, the plus solitons discussed in the next paragraph.

On the brighter side,  DQPTs with interesting features emerge in soliton dynamics. These special particle-like solutions are produced
from  eigenstates of the Fermi gas [normal solitons] or unstable stationary states of the superfluid [anomalous solitons]. The latter solitons are of two types, which we dubbed plus and minus solitons.  Remarkably, it turns out that DQPTs count the total number of `particles' [solitons]  for  normal multi-soliton dynamics and the number of minus particles for the anomalous multi-solitons. However, like solitons themselves  DQPTs  are  transient and their nice properties      reflect the  quantized nature of solitons just  as  other quantities, such as total energy, momentum etc., generally do.  And vice versa, from the point of view of DQPTs normal solitons, for example, are just a subtype of states with maximal discontinuity
in $n(\xi)$ at the Fermi energy.
Further,  plus solitons
possess no DQPTs despite the fact that their many-body wavefunction is similarly distinct from the BCS ground state at all times and, in particular, $n(\xi)$ is discontinuous.

 Our study of quenched  $p_x+ip_y$ fermionic superfluids revealed a major deficiency in the notion of DQPTs 
 when applied to time-dependent Hamiltonians that do not conserve the total number of particles, such as  mean field BCS Hamiltonians. In particular, the Loschmidt echo is not invariant with respect to global $U(1)$ rotations of the superfluid phase and there are no DQPTs even for quenches across the quantum critical point. We were able to repair this deficiency by introducing the more general  notion of the grand canonical Loschmidt echo. In our study of the
 $s$-wave BCS dynamics, this issue was masked by a particle-hole symmetric choice of  initial conditions for which the grand canonical and the usual notions of the Loschmidt echo coincide. DQPTs do emerge in the quench dynamics of the $p_x+ip_y$ superfluids when using the grand canonical Loschmidt echo. Furthermore, one can tell by their number   whether or not Majorana edge modes have appeared or disappeared as a result of the quench and whether the quench was across the quantum critical point or not.
 
 An interesting open problem is to investigate the Loschmidt amplitude  starting with the  four-fermion  $s$- or $p$-wave BCS Hamiltonian [Eqs.~\re{H} or~\re{fullpwave}] using, e.g., the path integral approach. Then, with an appropriate Hubbard-Stratonovich transformation    one should be able to see how the grand canonical Loschmidt amplitude ${\cal G}_\mu(t)=\langle\Psi_i | \Psi_\mu(t)\rangle$ emerges 
 naturally as the correct description rather than the overlap $\langle\Psi_i | \Psi_\mathrm{BCS}(t)\rangle$
 of the initial state with the time evolved BCS wavefunction.  In the thermodynamic limit, we expect  ${\cal G}_\mu(t)$ to reproduce the
 exact Loschmidt amplitude for the  four-fermion BCS Hamiltonians.

 While the Loschmidt echo cannot tell us about the nonequilibrium dynamics of an interacting system, there are other applications for this quantity where it can be useful. For example, we saw in our analysis of interaction quenches in the $p_x+ip_y$ superfluid that DQPTs carry information about the difference in \textit{equilibrium} properties of final and initial Hamiltonians. In this regard, it is worthwhile to emphasize the connection between the Loschmidt amplitude and the partition function of the system~\cite{HeylPolKeh}.   Consider a time-independent Hamiltonian $\hat H$, such as the four-fermion BCS Hamiltonian before mean field decoupling. Expanding the initial state
in terms of the eigenstates of $\hat H$, $\left| \Psi_i \right> = \sum_n c_n \left|\Psi_n \right>$, we deduce that
\beg
\mathcal{G}(t) = \sum_n \left| c_n \right|^2
e^{-i E_n t},
\en
where $E_n$ are the eigenvalues of the Hamiltonian. For $|c_n|=1$ this is the canonical partition function analytically continued to imaginary temperature $T=i t^{-1}$.  The grand canonical Loschmidt amplitude similarly corresponds to the grand canonical partition function, which justifies the terminology  introduced in this paper.

 The Loschmidt echo is in turn closely related to the spectral form factor~\cite{HalimehYegGurar}. Indeed,
\beg 
\mathcal{L}(t) = \sum_{nm} \left| c_n \right|^2
\left| c_m \right|^2 e^{-i t (E_n-E_m)},
\en
 which coincides with the spectral form factor ${\cal S}(t)$ when ${|c_n|=1}$. The Fourier transform  $\widetilde{{\cal S}}(\omega)$ of  the spectral form factor with respect to time $t$ is the correlation between the density of many-body states at energies separated by $\omega$. A non-analyticity in $\mathcal{L}(t)$ i.e., a DQPT, at $t=t_\mathrm{DQPT}$ is therefore a reflection of strong correlation between many-body energy levels separated by $\omega_0 \sim t_\mathrm{DQPT}^{-1}$.
 This also clarifies the role of $c_n$, i.e., of the initial state. A particular choice of the initial state can suppress or emphasize the correlations in certain parts of the spectrum.   It would be interesting to see if this connection between the Loschmidt echo and  spectral form factor can help reveal the true meaning and significance of various DQPTs we encountered in this work.


\acknowledgements{ Research at the University of Maryland was supported by US-ARO Contract No.W911NF1310172, NSF DMR-2037158 (V.G. and C.R.). The collaborative work between the University of Colorado and the University of Maryland was supported by the Simons Collaboration on Ultra-Quantum Matter, which is a grant from the Simons Foundation (651440, V.G. and V.G.).
 A. Z. is partially supported by Grant No.\ 2018058 from the United States-Israel Binational Science Foundation (BSF)}
\bibliography{bib}
\end{document}